\documentclass[a4paper,12pt]{article}
\usepackage[utf8]{inputenc}
\usepackage[english,russian]{babel}
\usepackage[]{fontenc}
\usepackage{hyperref,amsfonts,amssymb,amsmath}
\usepackage{graphicx}
\usepackage{mathrsfs}
\usepackage{pb-diagram}

  \def\Alambdai{{A}_{\lambda,\cdot}}
 \def\C{\mathbb{C}}
 \def\Z{\mathbb{Z}}
 \def\E{\mathcal{E}}
 \def\Fo{\mathscr{F}}
 \def\Heis{{\mathcal U}}
 \def\H{\mathscr{H}}
 
 \def\gl{\mathfrak{gl}}
 \def\Lambdai{{\Lambda}_{\lambda,\cdot}}
 
 \def\MNbar{\bar{M}_N}
 \def\Mlambda{\bar{M}_{\lambda}}
 \def\Mlambdai{\bar{M}_{\lambda,i}}
  \def\olambda{{\lambda}}
  \def\ooplus{\operatorname{\oplus}}
  \def\tV{\tilde{V}}
  \def\Q{Q}
  \def\V{V}
  \def\Yb{}
  \def\Ya{}
  \def\ZZ{{\mathbb Z}}

\usepackage[usenames,dvipsnames]{color}
\usepackage{ulem}

\begin{document}

\def\rf#1{(\ref{#1})}
\newtheorem{Def}{Definition}
\newtheorem{Th}{Theorem}
\newtheorem{Lemma}{Lemma}
\newtheorem{Lem}{Лемма}
\newtheorem{Stat}{Statement}
\newtheorem{Cons}{Corollary}
\newtheorem{Conjecture}{Conjecture}
\newtheorem{Prop}{Proposition}
\newtheorem{Remark}{Remark}
\newenvironment{Proof}{\par\noindent{\bf Proof.}}{\hfill$\scriptstyle\blacksquare$}
\bigskip
\hfill{ITEP-TH-20/15}
\begin{center}
{\Large\bf On  Spin Calogero-Moser system at infinity}

\bigskip
{\bf S.M. Khoroshkin$^{\star\circ}$,  \, M.G. Matushko$^{\circ}$ and E.K. Sklyanin$^{\ast}$
}\medskip\\
$^\star${\it Institute for Theoretical and Experimental Physics, Moscow 117259, Russia;}\smallskip\\ 
$^\circ${\it National Research University Higher School of Economics, Myasnitskay Ulitsa, Moscow 101000, Russia;}\smallskip\\
$^\ast${\it  Department of Mathematics, University of York, York YO10 5DD, United Kingdom}
\end{center}
\selectlanguage{english}
\begin{abstract}\noindent
We present a construction of a new integrable model as an infinite limit of 
Calogero models of N particles with spin. It is implemented in the multicomponent Fock space. Explicit formulas for Dunkl operators, the Yangian
generators in the multicomponent Fock space are presented. The classical limit of the system is examined.
\end{abstract}

\section{Introduction}
The Calogero-Moser-Sutherland (CMS) model of $N$ bosonic particles with spins has two properties: integrability 
and the presence of an algebra of symmetries \cite{KK, CALOGERO}. The symmetries provide a representation of the Yangian $Y(\mathfrak{gl}_s)$. 
Both of these properties relate to the Dunkl operators: the integrals of motion may be expressed through its symmetic combinations, and the corresponding representation of the Yangian uses 
the same Dunkl operators and $N$ copies of the algebra $\mathfrak{gl}_s$. It is natural to develop  analogues of the CMS model for systems with
an infinite number of particles. 
Such systems were defined and studied   by  I. Andric, A. Jevicki and H. Levine \cite{AJ}, 
H. Awata,  Y. Matsuo, S. Odake, J. Shiraishi \cite{AMOS} and others in the framework of ``collective field theory'' in the spinless case and
by  H. Awata, Y. Matsuo and T. Yamamoto \cite{AMY} for  models with spins.

Effective and rigorous constructions of limits of quantum Calogero-Sutherland (CS) systems have attracted the attention of mathematicians for many years. Note first the fundamental research of D.Uglov \cite{Uglov}, where he defined and studied a fermionic inductive
limit of the CS system. His construction waited for more than 15 years for a further development until M. Nazarov and E.Sklyanin
suggested a precise construction of higher Hamiltonians  for an Uglov type limit of the scalar CS system using the Sekiguchi determinant 
and  the machinery of symmetric functions \cite{NazSk1}. In \cite{SerVes2} A.Veselov and A.Sergeev  suggested to define a bosonic
limit of the CS system as a projective limit of finite  models. Precise bosonic constructions of higher Hamiltonians in a Fock space 
were then presented by M.Nazarov and E.Sklyanin in \cite{NazSk} and by  A.Veselov and A.Sergeev in \cite{SerVes}. 
It was also shown in \cite{NazSk} that the classical limit of the infinite system can be identified with the Benjamin-Ono 
hierarchy  whose Lax presentation can be derived from the suggested description of the quantum system. Moreover,  the equivariant family of Heckman--Dunkl operators can be regarded as a quantum $L$-operator for the CS system.

This paper can be regarded as a extension of the latter ideas to spin CMS models.
Our motivation to perform such an exercise was to reproduce an expression for the first Hamiltonian given 
in \cite{AMY} and to find a regular way to obtain similar expressions for the Yangian generators and higher 
Hamiltonians of the systems. We  first perform a translation of the finite spin CMS model to the language of polysymmetric functions and then realize the  
projective type limit \Yb of the spin CMS model in the multicomponent Fock space. Here we present  explicit formulas for the Dunkl operators, and the $L$-operator, which are then used for construction
of  the Yangian generators and the commuting Hamiltonians. \Ya Note
 that almost the same formulas are valid in the finite case as well. In particular, we reproduce  an expression for 
the first Hamiltonian  from \cite{AMY}. The answers are given now by integral operators with kernels composed of products of vertex operators.

We study the classical limit of the obtained system by sending the coupling constant to infinity. 
In the resulting model the Hamiltonians are constructed, and the equations of motion are presented. 
The classical system, which may be regarded as a multicomponent version of the Benjamin-Ono hierarchy, is integrable. 
It possesses an infinite number of commuting integrals of motion. 
Moreover, it is possible to express the equations of motion in Lax form: we present an explicit expression for
the Lax pair.

\section{CMS model and related algebraic structures}
\subsection{Calogero-Sutherland model}
Consider the quantum Calogero-Sutherland spin model of $N$ bosonic particles on the circle \cite{KK}. Its Hamiltonian is
\begin{equation*}
H^{CS}= -\sum_{i=1}^{N}\left(\frac{\partial}{\partial q_i}\right)^2+\sum_{i,j=1}^N\frac{\beta(\beta-1)}{\sin^2(q_i-q_j)}.
\end{equation*}
After conjugating by the function $|\prod_{i<j}\sin(q_i-q_j)|^\beta$ which represents the vacuum state, and passing to the exponential variables $x_i=e^{2\pi i q_i}$ we come to the Hamiltonian  
\begin{equation}\label{H}
H=\sum_{i=1}^{N}\left(x_i \frac{\partial}{\partial x_i}\right)^2+\beta\sum_{i<j}\frac{x_i+x_j}{x_i-x_j}\left(x_i\frac{\partial}{\partial x_i}-x_j\frac{\partial}{\partial x_j}\right)-2\beta\sum_{i<j}\frac{x_i x_j}{\left(x_i-x_j\right)^2}\left(1-K_{ij}\right),
\end{equation}
where the operator $K_{ij}$ permutes the variables $x_i$ and $x_j$.
The wave function of the Hamiltonian depends \Ya on positions $x_i$ of the particles and of their spins $\sigma(i)$,
 which  take values in the set $\left\{1,2,3,\dots , s\right\}$. \Yb
 We will consider polynomial wave functions invariant under the exchanges
\begin{equation}
\label{Prop}
\psi\left(\dots , x_i^{(a)},\dots , x_j^{(b)}\dots \right)=\psi\left(\dots , x_j^{(b)},\dots , x_i^{(a)},\dots \right),
\end{equation}
where the notation $x_i^{(a)}$ means that $i$-th particle has the position $x_i$ and spin $a=1,\dots,s$. 
The Dunkl operators
\begin{equation}\label{condit1}
d_i=x_i\frac{\partial}{\partial x_i}+\beta \sum_{j<i}\frac{x_j}{x_i-x_j}\left( 1 - K_{ij}\right)+\beta \sum_{i<j}\frac{x_i}{x_i-x_j}\left( 1 - K_{ij}\right)+\beta\left(i-1\right)
\end{equation} 
act on the space of wave functions and respect the condition (\ref{Prop}), see Section 2.2 for more details.
These operators commute
 $$
 [d_i,d_j]=0
 $$
 and give the expression of the Hamiltonian
 $$
 H =\sum_{i}\left(d_i^2 -\beta d_i\right),
 $$
 which is regarded as a linear operator on the space of functions with
property (\ref{Prop}).
  The higher commuting Hamiltonians, which act on the same space have the
form
\begin{equation}
\label{HH}
H_n =\sum_{i}d_i^n.
\end{equation}

\subsection{ Yangian and  degenerate affine Hecke algebra}

Denote by $M_{N}$ the space of vector-valued polynomials
\begin{equation}
M_{N}=\underbrace{(\mathbb{C}^s\otimes\mathbb{C}[z])\otimes\dots\otimes(\mathbb{C}^s\otimes\mathbb{C}[z])\otimes(\mathbb{C}^s\otimes\mathbb{C}[z])}_N.
\end{equation}
Then the space of the model can be identified with the space of $S_N$-invariants in $M_N$ denoted by $M_{N}^{S_N}$, where the symmetric group exchanges both the variables and vector components in the tensor product. The spaces of spins and of  coordinates are denoted by $\mathbb{C}^s$ and $\mathbb{C}[z]$, respectively.

More precisely, we fix a basis $\{e^1,e^2,\dots,e^s\}$ in the spin space $\mathbb{C}^s$. 
The operators $K_{ij}$ of permutation of the coordinates, $P_{ij}$ of permutation of the spins, and  $\sigma_{ij}=K_{ij}P_{ij}$ of the corresponding total action of the symmetric group $S_n$ can be expressed by the following formulas
$$
K_{ij}:\left(\dots\otimes(e^{a_i}\otimes x^{k_i})\otimes\dots\otimes(e^{a_j}\otimes x^{k_j})\otimes\dots\right)\to\left(\dots\otimes(e^{a_i}\otimes x^{k_j})\otimes\dots\otimes(e^{a_j}\otimes x^{k_i})\otimes\dots\right)
$$
$$
P_{ij}:\left(\dots\otimes(e^{a_i}\otimes x^{k_i})\otimes\dots\otimes(e^{a_j}\otimes x^{k_j})\otimes\dots\right)\to\left(\dots\otimes(e^{a_j}\otimes x^{k_i})\otimes\dots\otimes(e^{a_i}\otimes x^{k_j})\otimes\dots\right)
$$
$$
\sigma_{ij}:\left(\dots\otimes(e^{a_i}\otimes x^{k_i})\otimes\dots\otimes(e^{a_j}\otimes x^{k_j})\otimes\dots\right)\to\left(\dots\otimes(e^{a_j}\otimes x^{k_j})\otimes\dots\otimes(e^{a_i}\otimes x^{k_i})\otimes\dots\right)
$$
Any element in $M_{N}^{S_N}$ is invariant under the action of the symmetric group $\sigma_{ij}\psi=\psi$ by definition. Hence,
$$
K_{ij}\psi=P_{ij}\psi,\ \ \psi\in M_{N}^{S_N}.
$$
Next we describe the representation of the degenerate affine Hecke algebra and Lie algebra 
$\mathfrak{gl}_s$ in the space $M_N$. The Heckman--Dunkl operators $D_i^{(N)}:M_N\rightarrow M_N$ are given by
\begin{equation} \label{Dunkl}
D_{i}^{(N)}=x_i\frac{\partial}{\partial x_i}+\beta\sum_{j\neq i}\frac{x_i}{x_i-x_j}\left( 1 - K_{ij}\right).
\end{equation}
These operators do not change spins and satisfy the relations
\begin{align}\label{1}
K_{ij}D_i^{(N)}&=D_j^{(N)} K_{ij}, \\
\label{2}
[D_i^{(N)},D_j^{(N)}]&=\beta (D_j^{(N)}-D_i^{(N)})K_{ij},
\end{align}
which coincide with the relations of the degenerate affine Hecke algebra after the renormalization 
${D_i^{(N)}}=\beta\tilde{D}_i$. 
One can use another set of commuting elements $d_i$ of the degenerate affine Hecke algebra:
\begin{equation}\label{Dd}
d_i=D_i^{(N)}+\beta\sum_{j<i}K_{ij},
\end{equation}
so that
$$
[d_i,d_j]=0,\qquad 
K_{i,i+1}d_i=d_{i+1}K_{i,i+1}-\beta.
$$
The action of elements $d_i$ in the space $M_N$ is given by equation \rf{condit1}. The center of the degenerate affine Hecke algebra is represented by  symmetric functions of $d_i$. Moreover,  symmetric polynomials of $d_i$ represent integrals of motion of the quantum Calogero Sutherland model (\ref{HH}). 

The spin Calogero-Sutherland model admits Yangian symmetry, namely there is an action of the Yangian on 
the space $M_{N}^{S_N}$ which commutes with the Hamiltonians (\ref{HH}). Further we
recall the definition of the Yangian and describe its action on space $M_{N}^{S_N}$.
Let 
$$
R_{12}(u-v)=1-\frac{1}{u-v}\sum_{a,b}E^{ab}\otimes E^{ba} \in \mathrm{End}\,(\mathbb{C}^s\otimes \mathbb{C}^s),
$$
where $E^{ab}\in \mathrm{End}\,\mathbb{C}^s$ is the matrix unit, $E^{ab}(e^c)=\delta_{bc}e^a$ for basic vectors $e^c\in \mathbb{C}^s$.
By definition, the Yangian $Y(\mathfrak{gl}_s)$ is a unital associative algebra over $\mathbb{C}$ 
with generators $t^{ab}_{(i)}$, $a,b=1,...,s$, $i=0,1,...$ subject to the relations encoded in the Yang-Baxter equation
\begin{equation}\label{Yangrel}
R_{12}(u-v)t_1(u)t_2(v)=t_2(v)t_1(u)R_{12}(u-v),
\end{equation}
{\it where} $ t_1(u)=t(u)\otimes\text{Id}$, \  $ t_2(u)=\text{Id}\otimes t(u)$, {\it and}
$$
\qquad   
 t(u)=\sum_{a,b}t^{ab}(u)\otimes E^{ab}\in Y(\mathfrak{gl}_s)\otimes \mathrm{End}\,\left(\mathbb{C}^s\right)\,,
\qquad t^{ab}(u)=\delta^{ab}+\sum_{i=0}^{\infty}t^{ab}_{(i)}\frac{1}{u^{i+1}}\,.$$
Equivalently, the relations (\ref{Yangrel}) are
$$
\left[t^{ab}(u),t^{cd}(v)\right]=\frac{t^{cb}(u)t^{ad}(v)-t^{cb}(v)t^{ad}(u)}{u-v}.
$$
We have an action of $N$ copies of the Lie algebra $\mathfrak{gl}_s$ on $M_N$:
$$
E_{i}^{ab}\Big(\dots\otimes\underbrace{(e^{c}\otimes x^{k})}_i\otimes\dots\Big)=\delta_{bc}\Big(\dots\otimes\underbrace{(e^{a}\otimes x^{k})}_i\otimes\dots\Big).
$$
The global action of $\mathfrak{gl}_s$ on $M_N$ is denoted by $\mathbb{E}^{ab}=\sum_i E_i^{ab}$.

We have a representation of the Yangian $Y(\mathfrak{gl}_s)$ on the space $M_N^{S_N}$. We use the additive version of the Drinfeld functor to define the Yangian action \cite{Arakawa}:
\begin{equation}\label{repYan}
t^{ab}(u)= \delta_{ab}+\beta\sum_i \frac{E^{ab}_{i}}{\beta u+D_i^{(N)}} 
\end{equation}
on the space $M_N^{S_N}$. \Ya The following result is well known  \cite{B}: 
\medskip

\noindent \Ya {\bf Theorem} {\it The representation (\ref{repYan}) of Yangian $Y(\mathfrak{gl}_s)$  
commutes with the Hamiltonians of the CS system. The algebra of higher Hamiltonians
coincides with the center of the Yangian, which is generated by the coefficients of the quantum determinant
}
\Yb \begin{equation} 
q\det t(u)= \sum_{\sigma\in S_m}(-1)^{sgn(\sigma)}
t^{\sigma(1),1}(u)t^{\sigma(2),2}(u-1) ... t^{\sigma(m),m}(u-m+1).
\end{equation}
\section{CS system in the limit of infinitely many particles}
\subsection{Review of the scalar system}
\Ya Here we review recent results on the scalar CS system \cite{NazSk,SerVes} mainly  following the  approach of \cite{NazSk}. 
The main idea is to regard the equivariant Heckman-Dunkl operators as a quantum L-operator acting on the space of polynomial functions
of one  variable with coefficients being symmetric polynomials of the remaining $N-1$ variables.
  
In the scalar CS model the action of the Hamiltonian (\ref{H}) on symmetric functions can be reduced to
\begin{equation}\label{Hws}
H=\sum_{i=1}^{N}\left(x_i \frac{\partial}{\partial x_i}\right)^2+\beta\sum_{i<j}\frac{x_i+x_j}{x_i-x_j}\left(x_i\frac{\partial}{\partial x_i}-x_j\frac{\partial}{\partial x_j}\right).
\end{equation}
The corresponding Heckman--Dunkl operators $D_i^{(N)}:\C[x_1,\dots,x_N]\rightarrow \C[x_1,\dots,x_N] $ are defined by 
\rf{Dunkl}. Clearly, symmetric functions of $D_i^{(N)}$ preserve the ring of symmetric polynomials 
$\Lambda_N=\C[x_1,\dots,x_N]^{S_N}$.
This algebra is generated by the Newton polynomials $p_k^N=x_1^k+\dots +x_N^k$,  $k\in 0,1,\dots N$ \Ya
(sometimes we omit the upper index $N$ and simply write $p_k$). 
The Dunkl operator $D_i^{(N)}$ itself preserves the symmetry involving all variables other than $x_i$ and therefore 
it acts on the space $\Lambda_{N,i}$ of functions symmetric in all variables except $x_i$: \Yb
\begin{equation}\label{fun}
F(x_i,\{p_n\})
\in \Lambda_{N,i}\backsimeq\C[x_i]\otimes\C[x_1,\dots x_{i-1},x_{i+1},\dots x_N]^{S_{N-1}}.
\end{equation}
In the following 
 we use the notation
\begin{equation}\label{Vtil}
	\tilde{V}(z)=\exp\left(\sum_{n>0}z^n\frac{\partial}{\partial p_n}\right).
\end{equation}
for the linear map, which 
 changes each occurrence of a Newton sum $p_k^N$ by $p_k^{N-1}+z^k$.

Let $F\in\Lambda_N$ be a symmetric function of $\{p_n\}$ depending on $N$ variables.
\begin{Lemma} \label{lemma1}
	The natural embedding $\iota_{N,i}:\Lambda_N\rightarrow\Lambda_{N,i}$ is given by 
	\begin{equation}\label{4}
		\iota_{N,i}(F)=\tilde{V}(x_i)F.
	\end{equation}
	\end{Lemma}
	Here $\tilde{V}(x_i)F$ is a function of $x_i$ and $\{p_n\}$ depending on $(N-1)$ variables.
 \begin{Proof}
 \Ya The embedding $\iota_{N,i}$ can be regarded as the presentation of a symmetric function $F$
 by a polynomial in $x_i$ with coefficients being symmetric functions of the remaining variables: \Yb
$$
F(\{p_n\})=F_0(x_1,\dots x_{i-1},x_{i+1},\dots x_{N})+F_1(x_1,\dots x_{i-1},x_{i+1},\dots x_{N})x_i+
$$
$$
F_2(x_1,\dots x_{i-1},x_{i+1},\dots x_{N})x_i^2+\dots.
$$
This expansion can be obtained by means of a substitution
\begin{equation}\label{pnn}
p_n^N \rightarrow p_n^{N-1}+x_i^n
\end{equation}
which in its turn can be obtained by applying  the vertex operator (\ref{Vtil}) due to the Taylor formula
$$
f(z+t)=\exp(t\frac{\partial}{\partial z})=f(z)+f'(z)t+\frac{1}{2}f''(z)t^2+\dots
$$
which yields a finite sum for polynomials. Observe that the formula (\ref{4}) is correct for any expression of the symmetric function in terms of Newton polynomials $p_n$ irrespective of their dependencies.
\end{Proof}

Let $\phi^\pm (\xi)$ be the following power series in $\xi^{\pm 1}$:
\begin{equation}
\phi^+(\xi)=\sum_{n=1}^{\infty}\xi^n \frac{n\partial}{\partial p_n},\qquad 
\phi^-(\xi)=\sum_{n=0}^{\infty}\frac{p_n}{\xi^n},
\end{equation}
where $p_0=N$. We also use the notation
\begin{equation}\label{Vtil'}
	\tilde{V}'(z)=\exp\left(\sum_{n>0}-z^n\frac{\partial}{\partial p_n}\right).
\end{equation}
By definition \Ya the \Yb operator $\tilde{V}'(z)$ changes each occurrence of the formal variable $p_k^{N-1}$ 
by the difference $p_k^{N}-z^k$. The operator  $\tilde{V}'(x_i)$ maps the space  $\Lambda_{N,i}$ to $\Lambda_N\otimes \C[x_i]$.
 Note that 
 \begin{equation}\label{VV'}
   \tilde{V}'(z)\tilde{V}(z)F=F \qquad\forall\  F\in\Lambda_N.
 \end{equation}

 Define the operator $\E:\Lambda_{N,i}\rightarrow \Lambda_N$ of summation over $i$. For any $F(x_1,..., x_{i-1};z;x_{i+1},
 ..., x_N)$ $\in \Lambda_{N,i}$:
 \begin{equation*}
 \begin{split}
 \E F(x_1,...,x_N)=
 &F(x_i,..., x_{i-1};x_1;x_{i+1},
 ..., x_N)+F(x_1,x_i,..., x_{i-1};x_2;x_{i+1},
 ..., x_N)+...\\ +&F(x_1,..., x_{i-1};x_N;x_{i+1},
 ..., x_i).
 \end{split}
 \end{equation*}
 \begin{Lemma}\label{lemma2}
 	Let $F(x_i;\{p_n\})\in \Lambda_{N,i}$. Then 
 \begin{equation*}
 	\E F(\{p_n\})=\oint \frac{d\xi}{\xi}\phi^-(\xi)\left(\tilde{V}'(\xi)F\right)(\xi;\{p_n\}).
 \end{equation*}
 \end{Lemma}
 Here on the RHS the function $F(\xi;\{p_n\})$ depends on symmetric functions  of $(N-1)$ variables,
 while $F(\{p_n\})$ on the LHS  and $\tilde{V}'(\xi)F$ on the RHS both depend on symmetric functions on $N$ variables.  
\begin{Lemma}\label{lemma3}
The action of the Dunkl operator $D_i^{(N)}$ on functions $F(x_i,\{p_n\})\in \Lambda_{N,i}$  can be expressed by the following formula:
\begin{equation}\label{De}
D_i^{(N)}(F(x_i,\{p_n\}))=x_i\frac{\partial}{\partial x_i}F(x_i,\{p_n\})+
\beta x_i\oint\frac{d\xi}{\xi^2}\frac{\phi^-(\xi)-1}{1-\frac{x_i}{\xi}}\left(\tilde{V}'(\xi)\tilde{V}(x_i)F\right)(\xi,\{p_n\}),
\end{equation}
\end{Lemma}
The integral on the right hand side counts the residue at infinity:
$$
\oint f(\xi)d\xi=f_{-1}\ \text{for} \ f(\xi)= \sum_i f_i\xi^i.
$$
\begin{Proof}
In detail, the Dunkl operator $D_i^{(N)}$ transforms the space $\Lambda_{N,i}$ of functions with chosen variable $x_i$ into itself:
$$
D_i^{(N)}(F(x_i,\{p_n\}))=x_i\frac{\partial}{\partial x_i}F(x_i,\{p_n\})+\beta\sum_{j\neq i}\frac{x_i}{x_i-x_j}(1 - K_{ij})F(x_i,\{p_n\})=
$$
$$
=x_i\frac{\partial}{\partial x_i}F(x_i,\{p_n\})+\beta\sum_{j\neq i}\frac{x_i}{x_i-x_j}\left(\left(\tilde{V}(x_j)F\right)(x_i,\{p_n\})-\left(\tilde{V}(x_i)F\right)(x_j,\{p_n\})\right)=
$$
In each occurrence of $F(x_i;\{p_n\})$ we regard $\{p_n\}$ as symmetric functions of $(N-1)$ variables, 
while in $\left(\tilde{V}(x_j)F\right)(x_i,\{p_n\})$ $\{p_n\}$  depend on $(N-2)$ variables 
(all except $x_i$ and $x_j$). Using the absence of singularities on the diagonals $x_i=x_j$ for Dunkl operators,
we first present each fraction in the series as a function of $x_i/x_j$, then replace them by Cauchy integrals, to get: 
$$
D_i^{(N)}(F(x_i,\!\{p_n\}))=x_i\frac{\partial}{\partial x_i}F(x_i,\!\{p_n\})-$$
\begin{align*} &
\beta\sum_{j\neq i}\frac{\frac{x_i}{x_j}}{1-\frac{x_i}{x_j}}\left(\left(\tilde{V}(x_j)F\right)(x_i,\{p_n\})
-\left(\tilde{V}(x_i)F\right)(x_j,\{p_n\})\right)=
x_i\frac{\partial}{\partial x_i}F(x_i,\{p_n\})+\\
&\beta x_i\oint\frac{d\xi}{\xi^2}\frac{\phi^-(\xi)-1}{1-\frac{x_i}{\xi}}\left(\tilde{V}'(\xi)\tilde{V}(x_i)
F\right)(\xi,\{p_n\})-
\beta x_i\oint\frac{d\xi}{\xi^2}\frac{\phi^-(\xi)-1}{1-\frac{x_i}{\xi}}\left(\tilde{V}'(\xi)\tilde{V}(\xi)F\right)(x_i,\{p_n\}).
\end{align*}
In the last summand the vertex operators  cancel each other due to (\ref{VV'}), and the corresponding integral
vanishes since it contains $\xi$ only in negative powers. We  then obtain formula (\ref{De}). \Yb
\end{Proof}

\Ya The ring $\Lambda_\infty$ of symmetric functions with infinite number of variables is defined as the projective limit $\Lambda_\infty=\underleftarrow{\lim}\Lambda_N$ 
with respect to the projection $\Lambda_{N+1}\rightarrow\Lambda_N$: \Yb
$$
f\left(x_1,x_2,\dots,x_N,x_{N+1}\right)\rightarrow f\left(x_1,x_2,\dots,x_N,0\right).
$$
An element of $\Lambda_\infty$ can be represented by a sequence of symmetric functions:
\begin{equation}\label{ss}
f_1(x_1),f_2(x_1,x_2),\dots, f_N(x_1,x_2,\dots,x_N),\dots,
\end{equation}
that stabilizes $f_{N+1}\left(x_1,x_2,\dots,x_N,0\right)= f_N\left(x_1,x_2,\dots,x_N\right)$.

The ring $\Lambda_N$ is generated by Newton power sums $p_k^{(N)}\left(x_1,x_2,\dots,x_N\right)=
\sum_{i=1}^N x_i^k$  $(n\leqslant N)$. The Newton polynomials satisfy the stability condition (\ref{ss}) and thus correctly define an element $p_k\in \Lambda_{\infty}$ that can be presented as a series
$p_k=\sum_i x_i^k$. 
\Ya The elements $p_k$, $k=0,1,\ldots$ freely generate the ring $\Lambda_\infty$. Note that we add to $\Lambda_\infty$ 
 the formal variable $p_0$.
The canonical projection $ \Lambda_\infty \rightarrow \Lambda_N $ can be desribed by the relation:
\begin{equation}\label{pro}
 \Lambda_\infty \rightarrow \Lambda_N :\ \  p_k\rightarrow p_k^{(N)}=\sum_{i=1}^N x_i^k,\ \ p_0\rightarrow N.
\end{equation}

As is well known, $\Lambda_\infty$ is a realization of the bosonic Fock space, 
which is an irreducible representation of the Heisenberg algebra $\Heis$, generated by elements $\alpha_n$, $n\in\Z$ subject to relations $$\left[\alpha_n,\alpha_m\right]=n\delta _{n,-m}\,,$$
$$
\alpha_n\rightarrow n\frac{\partial}{\partial p_n},\ \ \alpha_{-n}\rightarrow p_n,\ n>0,\qquad \alpha_0\to p_0.
$$
We identify $\Lambda_\infty$ with the Fock space $\Fo$.

We use \Ya the \Yb operator $V(z):\Fo\otimes\C[z]\rightarrow\Fo\otimes\C[z]$
\begin{equation}\label{V}
V(z)=\exp\left(\sum_{n \geqslant0}z^n\frac{\partial}{\partial p_n}\right).
\end{equation}
Define an operator $D:\Fo\otimes\C[z]\rightarrow\Fo\otimes\C[z]$
\begin{equation}\label{DE}
D(F(z,\{p_n\}))=z\frac{\partial}{\partial z}F(z,\{p_n\})+
\beta z\oint\frac{d\xi}{\xi^2}\frac{\phi^-(\xi)}{1-\frac{z}{\xi}}V^{-1}(\xi)V(x_i)F(\xi,\{p_n\}),
\end{equation}
There is a map $\pi_i^{(N)}:\Fo\otimes\C[z]\rightarrow\Lambda_{N,i}$:
$$
p_k\rightarrow \sum_{j\neq i}^N x_j^k, \ z\rightarrow x_i.
$$
By (\ref{De}) and (\ref{DE}), we get the following commutative diagram:
\begin{equation}\label{diag}
\begin{diagram}
\node{\Fo\otimes\C[z]}
\arrow[2]{e,t}{\pi_i^{(N)}}
\arrow{s,l}{D}
\node[2]{\Lambda_{N,i}} \arrow{s,r}{D_i^{(N)}} \\
\node{\Fo\otimes\C[z]}
\arrow[2]{e,b}{\pi_i^{(N)}}
\node[2]{\Lambda_{N,i}} 
\end{diagram}
\end{equation}

There is a difference between the definitions (\ref{Vtil}) and (\ref{V}) for the vertex operators $\tilde{V}(x)$ and
$V(z)$ in the zero mode $\frac{\partial}{\partial p_0}$. This difference is compensated by the term $\phi^-(\xi)-1$ in (\ref{De}). The presence of $\frac{\partial}{\partial p_0}$ in $V(z)$ is responsible for a decrease in the number of particles in the zero-mode $p_0$.
The commutativity of the diagram (\ref{diag}) implies 
\begin{Prop} 
The operator $D$ (\ref{DE}) is a projective limit of Dunkl operators $D_i^{(N)}$\!, see \cite{NazSk} and \cite{SerVes}.
\end{Prop}
 Define the linear map $\E:\Fo\otimes \C[z]\to\Fo$ as
\begin{equation*}
\E F(z,\{p_n\})=\oint \frac{d\xi}{\xi}\phi^-(\xi)V^{-1}(\xi)F(\xi,\{p_n\}).
\end{equation*}
Then the  operators  $\H_k=\E D^k\iota:\Fo\to\Fo$, 
\begin{equation*}
\H_k:\Fo\xrightarrow{\iota}\Fo\otimes\C[z]\xrightarrow{D^k}\Fo\otimes\C[z]\xrightarrow{\E}\Fo,
\end{equation*}
generate a commutative family of Hamiltonians of the limiting system \cite{NazSk,SerVes}.

As an example let us calculate the Hamiltonian $\H_2$:
\begin{equation*}
\H_2\, F(p)=\oint_{\xi} \frac{d\xi}{\xi}\phi^-(\xi)V^{-1}(\xi)\left(\xi \frac{\partial}{\partial \xi}\right)^2 V(\xi)F(p)+
\end{equation*}
$$
+\beta\oint_{\eta<\xi}\frac{d\xi d\eta}{\xi^2}\phi^-(\eta)V^{-1}(\eta)\frac{\phi^-(\xi)}{1-\frac{\eta}{\xi}}V(\eta)V^{-1}(\xi)\left(\xi\frac{\partial}{\partial \xi}\right)V(\xi)F(p)=
$$
$$
=\oint_{\xi} \frac{d\xi}{\xi}\phi^-(\xi)\left(\sum_{k>0,n>0}k n \xi^{k+n}\frac{\partial}{\partial p_k}\frac{\partial}{\partial p_n}+\sum_{n>0}n^2 \xi^n \frac{\partial}{\partial p_n}\right)F(p)+
$$
$$
+\beta\oint\frac{d\xi d\eta}{\xi^2}\phi^-(\eta)\left(\frac{\phi^-(\xi)}{1-\frac{\eta}{\xi}}-\frac{1}{(1-\frac{\eta}{\xi})^2}\right)\sum_{n>0}\xi^n \frac{\partial}{\partial p_n}F(p).
$$
We have taken the second derivative in the first integral and used the commutator relations
$$[\phi^-(\xi),V(\eta)]=-\frac{V(\eta)}{(1-\frac{\eta}{\xi})}$$ in the second. Thus
$$
\H_2=\sum_{k>0,n>0}k n p_{k+n}\frac{\partial}{\partial p_k}\frac{\partial}{\partial p_n}+\sum_{n>0}n^2 p_n \frac{\partial}{\partial p_n}-\beta \sum_{n>0}n^2 p_n \frac{\partial}{\partial p_n}+\beta\sum_{k>0,n\geqslant 0}(k+n)p_k p_n\frac{\partial}{\partial p_{k+n}}.
$$

\subsection{The Spin CS system and  multicomponent Fock space}
In this section we present a generalization of the calculations in section 3.1 to the spin CS model. 
Let us split the difference part of the Dunkl operator into two sums
\begin{equation}
	D_{i}=x_i\frac{\partial}{\partial x_i}+\beta\sum_{j\neq i,\sigma(i)=\sigma(j)}\frac{x_i}{x_i-x_j}\left( 1 - K_{ij}\right)
	+\beta\sum_{j,\sigma(i)\neq\sigma(j)}\frac{x_i}{x_i-x_j}\left( 1 - K_{ij}\right).
\end{equation} 
The Dunkl operators act on the space $M_N$ and the symmetric functions of $D_{i}$ preserve the subspace of global $S_N$-invariants
$\MNbar:=M_N^{S_N}$, which we further identify with  the direct sum of spaces of symmetric functions by groups of variables. The space $M_N$ can be decomposed into the direct sum of weight spaces with respect to the diagonal action of the Cartan subalgebra of $\mathfrak{gl}_s$:
$$
M_N=\bigoplus_{|\lambda|=N}M_{\olambda},  
$$
\Ya where $\olambda=(\lambda_1,\lambda_2,\dots \lambda_s)$  is a collection of nonnegative integers,
$\sum_{c=1}^{s} \lambda_c=N$. \Yb
 In other words, $M_{\olambda}$  is the space of vector-valued polynomials generated by tensors of the form
\begin{equation}\label{monom}
	{\mathbf m}=\left(e_1\otimes z^{a_{11}}\cdots e_1\otimes z^{a_{1\lambda_1}}\right)\otimes 
	\left(e_2\otimes z^{a_{21}}\cdots e_2\otimes z^{a_{2\lambda_2}}\right)\otimes\cdots\otimes 
	\left(e_s\otimes z^{a_{s1}}\cdots e_s\otimes z^{a_{s\lambda_s}}\right)
\end{equation}
together with all their permutations. Denote by $\Mlambda:=M_{\olambda}^{S_N}$  the space of $S_N$-invariants in 
$M_{\olambda}$. Hence,
$$
\MNbar=\bigoplus_{|\lambda|=N}\Mlambda.
$$

Let $A_{\olambda}$ be the ring of polynomials in $N$ variables 
$(x_{11},x_{12},\dots x_{1\lambda_1}$, $x_{21},x_{22},\dots x_{2\lambda_2}\dots$ 
$x_{s1},x_{s2},\dots x_{s\lambda_s})$. 
Denote by $\Lambda_{\olambda}\in A_{\olambda}$ the ring of functions symmetric within each group, defined by the partition $\lambda$, of 
variables $(x_{11},x_{12},\dots x_{1\lambda_1})$, $(x_{21},x_{22},\dots 
x_{2\lambda_2})\dots (x_{s1},x_{s2},\dots x_{s\lambda_s})$.
Define a linear map $\gamma_\lambda:A_{\olambda}\rightarrow M_{\olambda}$. This map associates 
to each monomial $$
{x}={x_{11}}^{a_{11}}\cdots \ {x_{1\lambda_1}}^{a_{1\lambda_1}}\ \
\cdots \  \ {x_{s1}}^{a_{s1}}\cdots \ {x_{s\lambda_s}}^{a_{s\lambda_s}}$$
\Ya the  total symmetrization of the corresponding tensor ${\mathbf m}$, see (\ref{monom}), divided by the product
$\lambda_1!\cdots \lambda_s!$:
\begin{equation}\label{Shuf}
	\gamma_\lambda({x})=\frac{1}{\lambda_1!\cdots \lambda_s!}\sum_{\sigma\in S_N}\sigma(\mathbf{m}).
\end{equation}
The restriction of the map $\gamma_\lambda$ to $\Lambda_{\olambda}$ establishes an isomorphism of vector spaces
\begin{equation*}
 \gamma_\lambda:\Lambda_\lambda\backsimeq \Mlambda.
\end{equation*}
 Using this isomorphism we push forward the ring structure of $\Lambda_\lambda$ to $\Mlambda$.
 The ring $\Lambda_{\olambda}$ is generated by the (not indepent) elements
$$
p_{n,a}^{\olambda}=\sum_{i=1}^{\lambda_a}{x_{ai}}^n,
$$
 with $a=1,..,s$, $n=1,2,...$. Further we present elements of the ring $\Lambda_{\olambda}$ as polynomial
  functions $F(p)$ of the elements $p_{n,a}^{\lambda}$. 

For each $i=1,\ldots, N$ denote by $\Mlambdai$ the subspace of $M_\lambda$ which consists of tensors invariant with respect 
 to the symmetric group (isomorphic to $S_{N-1}$) of permutations of all tensor factors except the $i$-th one. 
 We have a natural inclusion (decomposition of a tensor 
 with respect to its $i$-th factor):
$$
\alpha_{\olambda,i}: \Mlambda\to\Mlambdai.
$$
 Denote by $\Alambdai$ and by 
 $\Lambdai$ the rings
 \begin{equation*}
 	 \Alambdai=\ooplus\limits_{c=1}^s \left(A_{\lambda-\epsilon_c}\otimes \C[z]\right),\qquad
  \Lambdai=\ooplus\limits_{c=1}^s \left(\Lambda_{\lambda-\epsilon_c}\otimes \C[z]\right).
 \end{equation*}
Here $\epsilon_1=(1,0,\dots,0)$, $\ldots \epsilon_s=(0,\dots,0,1)$. We present elements of the ring $\Lambdai$ as
 vector valued polynomial functions on elements $p_{n,a}^{\olambda}$ and $z$:
 $$\bar{F}(p,z)=\left(\begin{array}{c}
                       F_1(p,z) \\
                       ... \\
                       F_s(p,z)
                      \end{array}
\right).$$

 For each $i=1,\ldots, N$ we define a map $\gamma_{\lambda,i}:\Alambdai\to \Mlambdai$.  
  For the reasons of typographic convenience we first describe the map $\gamma_{\lambda,N}$. This map attaches to a monomial
  $x_c\in A_{\lambda-\epsilon_c}\otimes \C[z]$,
  $$x_c= 
  {x_{11}}^{a_{11}}\cdots \ {x_{1\lambda_1}}^{a_{1\lambda_1}}\ \ \cdots \ \ 
  {x_{c1}}^{a_{c1}}\cdots \  {x_{c,\lambda_c-1}}^{a_{c,\lambda_c-1}}
  \cdots \  \ {x_{s1}}^{a_{s1}}\cdots \ {x_{s\lambda_s}}^{a_{s\lambda_s}}\ \cdot z^d$$
 the symmetrization over the first $N-1$ tensor components of the tensor
 \begin{equation*}
 \begin{split}	
 	{\mathbf m}'=
 \left(e_1\otimes z^{a_{11}}\cdots e_1\!\otimes z^{a_{1\lambda_1}\!}\right) \cdots 
 \left(e_c\otimes z^{a_{c1}}\!\cdots e_c\otimes z^{a_{c,\lambda_c-1}}\!\right) \cdots  
 \left(e_s\otimes z^{a_{s1}}\!\cdots e_s\otimes z^{a_{s\lambda_s}}\!\right)\otimes\left( e_c\otimes z^d\right)\!,
 \end{split}
 \end{equation*}
  divided by the product
  $\lambda_1!\cdots(\lambda_c-1)!\cdots \lambda_s!$:
  \begin{equation*}
  	\gamma_{\lambda,N}({x_c})=\frac{1}{\lambda_1!\cdots (\lambda_c-1)!\cdots\lambda_s!}\sum_{\sigma\in S_{N-1}}\sigma(\mathbf{m}').
  \end{equation*}
  Next we set
  $$\gamma_{\lambda,i}=\sigma_{i,N}\cdot\gamma_{\lambda,N},$$
  where $\sigma_{i,j}$ is the transposition of the $i$-th and
 $j$-th tensor factors. By construction, the restriction of the map $\gamma_{\lambda,i}$ to $\Lambdai$ establishes an isomorphism of vector spaces
 \begin{equation*}
 	\gamma_{\lambda,i}:\Lambdai\approx\Mlambdai.
 	\end{equation*}
 	Restrictions of  $\gamma_{\lambda,i}$ to $\Lambdai$ form an equivariant family:
 	\begin{equation*}
 	\gamma_{\lambda,i}=\sigma_{i,j}\cdot\gamma_{\lambda,i}	.
 		\end{equation*}

For each $a=1,\ldots,s$ we define the  `vertex' operators
 $\tV_a(z):\,\Lambda_\lambda\to
\Lambda_{\lambda-\epsilon_a}\otimes\C[z]$ 
$$\tV_a(z)=\exp\left(\sum_{n > 0}z^n\frac{\partial}{\partial p^{\olambda}_{n,a}}\right).\qquad
$$ 
Thus we have the linear map
\begin{equation}\label{iotafin}
 \iota_{\olambda}:\Lambda_{\olambda}\rightarrow \Lambdai,\qquad \iota_{\olambda}: F(p)\rightarrow
 \tV_1(z,p)F(p)\oplus\dots\oplus \tV_s(z,p)F(p),
\end{equation}
that we also represent using the column-vector notation:
$$
\iota_{\olambda} F(p)=\begin{pmatrix}
\tV_1(z,p)F(p)\\
\dots\\
\tV_s(z,p)F(p)
\end{pmatrix}.
$$
\begin{Lemma} \label{lemma4} We have the equality
	$
	\gamma_{\lambda,i}\iota_{\olambda}=\alpha_{\olambda,i}\gamma_\lambda
	$.
\end{Lemma}
The commutative diagram below illustates   the statement from Lemma \ref{lemma4}. 
\begin{equation}\label{diag2}
	\begin{diagram}
		\node{\Lambda_{\olambda}}
		\arrow[2]{e,t}{\gamma_\lambda}
		\arrow{s,l}{\iota_{\olambda}}
		\node[2]{\Mlambda}
		\arrow{s,r}{\alpha_{\olambda,i}} \\
		\node{\Lambdai}
		\arrow[2]{e,b}{\gamma_{\lambda,i}}
		\node[2]{\Mlambdai}
	\end{diagram}.
\end{equation}
Next we translate the Dunkl operators to the language of polysymmetric functions.
 Let $\phi_a^\pm (\xi)$ be the power series in $\xi$ and in $\xi^{-1}$:
\begin{equation}
\phi_a^+(\xi)=\sum_{n=1}^{\infty}\xi^n n\frac{\partial}{\partial p^\lambda_{n,a}},\qquad
	\phi_a^-(\xi)=\sum_{n=0}^{\infty}
	\frac{p_{n,a}^\lambda}{\xi^n},
\end{equation}
where we set $p_{0,a}^\lambda=\lambda_a$. Let 
$\tV'_a(z)$ be the following operator from the polynomial ring of formal variables $p_{n,a}^\lambda$, $n>0$, $a=1,\ldots,s$ to $\Lambda_{\lambda+\epsilon_a}\otimes\C[z]$:
$$\tV'_a(z)=\exp\left(-\sum_{n > 0}z^n\frac{\partial}{\partial p^{\olambda}_{n,a}}\right).$$

Introduce the operator $D:\Lambdai\to \Lambdai$. It acts on a vector function $F(z,p)$ as a 
diagonal matrix, such that 
\begin{equation}\label{Dfin}
		D\left(F_a(z,p)\right)=z\frac{\partial}{\partial z}F_a(z,p)+
\beta\sum_b z\oint\frac{d\xi}{\xi^2}\frac{\phi_b^-(\xi)-\delta_{a,b}}{1-\frac{z}{\xi}}\tV'_b(\xi)\tV_b(z)F_a(\xi,p).
\end{equation}
\begin{Lemma}\label{lemma5} For any $i=1,...,N$ we have the equality 
$$\gamma_{\lambda,i}D=D_i\gamma_{\lambda,i}.$$
 \end{Lemma}
For $a,b=1,...,s$ define the linear map $\mathbb{E}^{ab}:\Lambdai\to\Lambda_{\lambda+\epsilon_a-\epsilon_b}$
by the relation
\begin{equation} \label{eabfin}
{\E}^{ab}F_c(z,p)=\delta_{bc}\oint_{\xi}\frac{d\xi}{\xi}\phi_a^-(\xi)\tV'_a(\xi)F_c(\xi,p).
	\end{equation}
	\begin{Lemma}\label{lemma6}
	 For any $F\in\Lambdai$ we have the relation
	 $$\gamma_\lambda {\E}^{ab}F= \sum_{i=1}^NE^{ab}_i\gamma_{\lambda,i}F.$$
	\end{Lemma}
In other words, ${\E}^{ab}$ is the operator of matrix averaging of a collection of equivariant tensors.

 The proof of lemmas \ref{lemma3}, \ref{lemma4}, \ref{lemma5} is analogous to that of lemmas \ref{lemma1}, \ref{lemma2}, \ref{lemma3}.
 
 Set $\Lambda^{s,N}=\ooplus_{|\mu=N|}\Lambda_\mu$. As a direct corollary of statements above we get 
 the Yangian action on $\Lambda^{s,N}$ translated from that on $\MNbar$.
 \begin{Prop}\label{cor1} The  Yangian generator $t^{ab}_{(k)}$ maps the space $\Lambda_\mu$ 
 to the space $\Lambda_{\mu +\epsilon_a-\epsilon_b}$ and 
 is given by the relation
 \begin{equation}\label{tabfin}
  t^{ab}_{(k)}F=\frac{(-1)^k}{\beta^k}{\E}^{ab}D^k\iota_\mu F\qquad\text{for any}\quad F\in\Lambda_\mu.
 \end{equation}
 where the maps $\iota_\mu$, $D$ and ${\E}^{ab}$ are given by the relations \rf{iotafin}, 
 \rf{Dfin}, \rf{eabfin}.
 \end{Prop}
 In terms of generating functions this action looks as follows:
 \begin{equation}\label{procedurefin}
t^{ab}(u)= \delta_{ab}+\beta{\E}^{ab}\left(\beta u+D\right)^{-1}\iota_\mu.
\end{equation}

Let $\Heis^s$ be the tensor product of $s$ copies of the Heisenberg algebra with generators $\alpha_{n,a}$ and $\Q_a^{\pm 1}$, where $n\in\Z$, $a=1,\ldots,s$ and  relations $$\left[\alpha_{n,a},\alpha_{m,b}\right]=n\delta_{a,b}\delta _{n,-m},\qquad Q_a\alpha_{n,b}Q_a^{-1}=\alpha_{n,b}+\delta_{n,0}\delta_{a,b}.$$
The multicomponent Fock space $\Fo^s$ is an irreducible representation of $\Heis^s$ in the space of polynomials $\C[p_{n,a},q_b^{\pm 1}\equiv e^{\pm\frac{\partial}{\partial p_{0,b}}}]$,
$ n\not=0,\ a,b=1,\ldots s$:
$$
\alpha_{n,a}\rightarrow n\frac{\partial}{\partial p_{n,a}},\ \ \alpha_{-n,a}\rightarrow p_{n,a},\ n>0,\qquad \alpha_{0,a}\to p_{0,a}-\lambda_a,\qquad \Q_a^{\pm 1}\mapsto q_a^{\pm 1},\qquad  \alpha_{0,a}\cdot 1=0,
$$
where $\lambda_1,\ldots, \lambda_s\in\C$ are arbitrary constants. 
For any $\nu\in\Z^s$ denote by $\Fo^s_\nu\in\Fo^s$ the subspace
$\C[p_{n,a}]q_1^{\nu_1}\cdots q_s^{\nu_s}$.
We have the decomposition
$$\Fo^s=\ooplus_{\nu\in\Z^s}\Fo^s_\nu.$$
 The subspace 
$\Fo^s_\nu$ is invariant with respect to all operators
$\alpha_{n,a}$ and
$\alpha_{0,a}v=\nu_a v$ for any $v\in \Fo^s_\nu$, or, in terms of $p_{0,a}$,
\begin{equation}
	\label{p0a}
	p_{0a} v=(\nu_a+\lambda_a)v \qquad\text{for any }
	\quad v\in\Fo^s_\nu.
	\end{equation}
	
	The multicomponent Fock space may be regarded as a projective limit of the spaces of polysymmetric functions in the following sense. Fix a natural number $N$ and a weight $\lambda=(\lambda_1,\cdots,\lambda_s)$ consisting of nonnegative integers such that $\sum_a \lambda_a=N$.
For any $\nu\in\Z^s$	define a map $\pi_{\lambda\nu}:\Fo^s_\nu\to\Lambda_{\lambda+\nu}$ by setting
	$$p_{n,a}\mapsto p_{n,a}^{\lambda+\nu},\quad n>0,\qquad\text{and} \quad p_{0,a}\mapsto \lambda_a+\nu_a.$$
 By definition, this map is nonzero only when all numbers $\nu_a+\lambda_a$ are nonzero. It is compatible with zero evaluation maps  $\tau_{\mu,a}:\Lambda_\mu\to
 \Lambda_{\mu-\epsilon_a}$,
 \begin{equation*}
 	\begin{split}&f(x_{11},...,x_{1\mu_1},\,...\,
 x_{a1},\,...\,,,x_{a\mu_{a}-1},x_{a\mu_a},\,...\,, x_{s1},\,...\,,x_{s\mu_1})\mapsto
\\ &f(x_{11},,\,...\,,x_{1\mu_1},\,...\,,
 x_{a1},\,...\,,x_{a\mu_{a}-1}\,,0\,,\,...\,, x_{s1},\,...\,,x_{s\mu_1}).
 \end{split}
 \end{equation*}
in the sense that
$$\pi_\lambda\cdot \tau_a=\pi_{\lambda-\epsilon_a},$$
where $\pi_\lambda=\ooplus_{\nu\in\Z^s}\pi_{\lambda_\nu}$,
$\tau_a=\ooplus_{\mu\in\Z_{\geq 0}^s}\tau_{\mu,a} $.

For any $a=1,...,s$ denote by $V_a(z):\Fo^s\rightarrow \Fo^s\otimes\C[z]$ the vertex operator
\begin{equation}
	V_a(z)=\exp\left(\sum_{n \geqslant 0}z^n\frac{\partial}{\partial p_{n,a}}\right).
\end{equation}
and use them for the embedding
 $\iota:\Fo^s\xrightarrow\iota\Fo^s\otimes\left(\C^s\otimes\C[z]\right)$, where
 	$ \iota_{} (F)=
 	\V_1(z)F\oplus\dots\oplus \V_s(z)F$, or, in vector notation
\begin{equation}\label{iotainf}
\iota F=\begin{pmatrix}
V_1(z)F\\
\ldots\\
V_s(z)F\\
\end{pmatrix}, \qquad F=F(p,q)\in \Fo^s.
\end{equation}
 We also use the fields
\begin{equation*}
	\phi_a^+(\xi)=\sum_{n=1}^{\infty}\xi^n n\frac{\partial}{\partial p_{n,a}},\qquad
	\phi_a^-(\xi)=\sum_{n=0}^{\infty}\frac{p_{n,a}}{\xi^n}.
\end{equation*}

Let us introduce the operator $D$. It acts on a vector function $F(z)\in\Fo^s\otimes\C[z]$ as a diagonal matrix $D:\Fo^s\otimes\left(\C^s\otimes\C[z]\right)\rightarrow \Fo^s\otimes\left(\C^s\otimes\C[z]\right)$ by
\begin{equation}\label{Dinf}
	D\left(F_a(z)\right)=z\frac{\partial}{\partial z}F_a(z)+\beta\sum_b z\oint\frac{d\xi}{\xi^2}\frac{\phi_b^-(\xi)}{1-\frac{z}{\xi}}V_b^{-1}(\xi)V_b(z)F_a(\xi).
\end{equation}
The operator
${\E}^{ab}:\Fo^s\otimes\left(\C^s\otimes\C[z]\right)\rightarrow \Fo^s$ of matrix averaging is now given by the relation
\begin{equation} \label{eabinf}
	{\E}^{ab}F_c(z)=\delta_{bc}\oint_{\xi}\frac{d\xi}{\xi}\phi_a^-(\xi)\V^{-1}_a(\xi)F_c(\xi).
\end{equation}
The maps $\iota, D$ and ${\E}^{ab}$ are compatible with finite
 projections $\pi_\lambda$ in the following sense. We formulate the compatibility conditions with the starting point at $\Fo^s_\nu\in\Fo^s$.
 \begin{Prop}\label{Prop3} For any $v\in\Fo^s$ and $u\in
\Fo^s\otimes\left(\C^s\otimes\C[z]\right)$
we have the following equalities:
   \begin{align*}&\left(\pi_\lambda\otimes 1 \right)
D(u)=D\left(\pi_\lambda\otimes 1 \right)(u),&&
   \iota_{\lambda+\nu}\pi_\lambda(v)=\left(\pi_\lambda\otimes 1\right)
   \iota(v),\\ & {\E}^{ab}\left(\pi_\lambda\otimes 1\right)(u)=
   \left(\pi_\lambda\otimes 1\right) {\E}^{ab}(u).
   \end{align*}
   \end{Prop}

As a consequence of proposition \ref{cor1} and proposition \ref{Prop3} we obtain the  representation $T_{k}^{ab}$ of the Yangian generators $t_{(k)}^{ab}$ in the multicomponent Fock space:

 \begin{Prop}\label{cor2} The assignement $t^{ab}_{(k)}\rightarrow T^{ab}_k$, where 
 	\begin{equation}\label{tabinf}
 	T^{ab}_{k}F=\frac{(-1)^k}{\beta^k}{\E}^{ab}D^k\iota F\qquad\text{for any}\quad F\in\Fo^s,
 	\end{equation}
 	defines a representation of the Yangian $Y(\gl_s)$ in the space $\Fo^s$.
 	 \end{Prop}
 	 	Here the maps $\iota$, $D$ and ${\E}^{ab}$ are given by the relations \rf{iotainf}, 
 	\rf{Dinf}, \rf{eabinf}.
 	In generating functions this action looks as follows:
 \begin{equation*}
T^{ab}(u)= \delta_{ab}+\beta{\E}^{ab}\left(\beta u+D\right)^{-1}\iota.
\end{equation*}
 The Yangian action can be illustrated by the following scheme: 
\begin{equation}\label{procedure}
	\beta^kT_{k}^{ab}:\ \Fo^s\xrightarrow{\iota}\Fo^s\otimes\left(\C^s\otimes\C[z]\right)\xrightarrow{(-D)^k}\Fo^s\otimes\left(\C^s\otimes\C[z]\right)\xrightarrow{{\E}^{ab}}\Fo^s.
\end{equation}

As an example we present corresponding integral expression for the zero and the first
generators of the Yangian  $Y(\gl_s)$:
\begin{align}\label{tab0}
T^{ab}_{0}= &\oint\frac{d\xi}{\xi}\phi_a^-(\xi)V_a^{-1}(\xi)V_b(\xi),\\
\label{tab1}\notag
T^{ab}_{1}=-\frac{1}{\beta}&\oint\frac{d\xi}{\xi}\phi_a^-(\xi)V_a^{-1}(\xi)\phi_b^+
(\xi)V_b(\xi)\\ &-\sum_c \oint\frac{d\xi d\eta}{\xi\eta}
\frac{\frac{\eta}{\xi}}{1-\frac{\eta}{\xi}}\phi_a^-(\eta)V_a^{-1}(\eta)\phi_c^-
(\xi)V_c(\eta)V_c^{-1}(\xi)V_b(\xi). 
\end{align}

\medskip

\subsection{Hamiltonians}
In this section we provide explicit expressions for the first few Hamiltonians constructed by means of the procedure (\ref{procedure}). The proof and calculation are presented in Appendix 4.
The Hamiltonians $H_{n}=\sum_i d_i^n$ can be expressed in terms of $D_i$ by means of (\ref{Dd}):
\begin{equation}
H_{1}=\sum_i D_i+\frac{\beta}{2}\sum_{a,b}\mathbb{E}^{ab}\mathbb{E}^{ba}-\frac{\beta}{2} s p_0,
\end{equation}
\begin{equation}
\begin{split}
H_{2}=\sum_i D_i^2+\beta\sum_{i,j,a,b}E^{ab}_iE^{ba}_jD_j-\beta s\sum_i D_i+
\\
+\frac{\beta^2}{3}\sum_{a,b,c}\mathbb{E}^{ab}\mathbb{E}^{bc}\mathbb{E}^{ca}
-\frac{2s\beta^2}{3}\sum_{a,b}\mathbb{E}^{ab}\mathbb{E}^{ba}+\beta^2\frac{2s^2+p_0-1}{6}p_0,
\end{split}
\end{equation}
 where $s$ and $p_0=\sum_a p_{0,a}$ are the numbers of spins and particles, respectively.
Using (\ref{cor2}) we can rewrite these formulas in terms of the Yangian generators $T^{ab}_k$:
\begin{equation}
	\frac{H_{1}}{\beta}=-\sum_a T^{aa}_{1}+\frac{1}{2}\sum_{a,b} T^{ab}_{0} T^{ba}_{0} -\frac{s}{2}\sum_a T^{aa}_{0},
\end{equation}
\begin{equation}
\begin{split}
	\frac{H_{2}}{\beta^2}=\sum_a T^{aa}_{2}-\sum_{a, b} T^{ab}_{0} T^{ba}_{1}+s\sum_a T^{aa}_{1}+\frac{1}{3}\sum_{a,b,c} T^{ab}_{0} T^{bc}_{0} T^{ca}_{0}
	\\-\frac{2s}{3}\sum_{a, b} T^{ab}_{0} T^{ba}_{0}+\frac{1}{6}\sum_{a,b} T^{aa}_{0}T^{bb}_{0}+\frac{2s^2-1}{6}\sum_a T^{aa}_{0}
	\end{split}
\end{equation}
On the other hand, the Hamiltonians can be obtained as elements of the $q$-determinant (Th1). Consider the representation of the $q$-determinant :
$$
\text{q-det}\,T(u)= \sum_{\sigma\in S_m}\!\!(-1)^{\text{sgn}(\sigma)}
T^{\sigma(1),1}(u)T^{\sigma(2),2}(u-1) \cdots T^{\sigma(m),m}(u-m+1)=1+\frac{\Delta_0}{u}+\frac{\Delta_1}{u^2}+\frac{\Delta_2}{u^3}+\dots,
$$
where
$$
T^{ab}(u)=\delta^{ab}+\frac{T^{ab}_0}{u}+\frac{T^{ab}_1}{u^2}+\frac{T^{ab}_2}{u^3}+\dots.
$$
The elements $\Delta_i$ can be expressed in terms of $T^{ab}_k$. Explicit expressions for $\Delta_0,\Delta_1,\Delta_2$ are presented in Appendix 3.
The Hamiltonians can be rewritten in the following form:
\begin{equation}
\frac{H_{1}}{\beta}=-\Delta_1+\frac{1}{2}\Delta_0^2-\frac{1}{2}\Delta_0,
\end{equation}
\begin{equation}
\frac{H_{2}}{\beta^2}=\Delta_2-\Delta_0\Delta_1+\Delta_1+\frac{1}{3}\Delta_0^3-\frac{1}{2}\Delta_0^2+\frac{1}{6}\Delta_0.
\end{equation}
\begin{Prop}
The first Hamiltonians in the multicomponent Fock space have the following form
\begin{equation}\label{h11}
\H_1=\sum_a\oint\frac{d\xi}{\xi}\phi_a^-(\xi)\phi_a^+(\xi)+\frac{\beta}{2}\left(p_{0}^2-p_{0}\right),
\end{equation}
\begin{align}\label{Haminf}
\notag
\H=\H_{2}-\beta\H_1=\sum_a\oint\frac{d\xi}{\xi}\phi_a^-(\xi)(\phi_a^+(\xi))^2+
(1-\beta)\sum_a\oint\frac{d\xi}{\xi}\phi_a^-(\xi)(\phi_a^+
(\xi))^\prime+
\\ \notag
-\beta\sum_a\oint\frac{d\xi}{\xi}\phi_a^-(\xi)\phi_a^+(\xi)+
\beta\sum_{a,b}\oint\frac{d\xi}{\xi}\phi_a^-(\xi)\phi_b^-(\xi)\phi_a^+(\xi)+
\\ 
\beta\sum_{a>b}\oint\frac{d\xi d\eta}{\xi\eta}\phi_a^-(\eta)\phi_b^-(\xi)
\sum_k k\left(\frac{\xi^k}{\eta^k}+
\frac{\eta^k}{\xi^k}\right) V_a^{-1}(\eta)V_b(\eta)V_b^{-1}(\xi)V_a(\xi).
\end{align}
\end{Prop}

\section{Classical limit}
In this section we investigate the classical limit of the Hamiltonian (\ref{Haminf}). In the spinless case it leads to the periodic Benjamin-Ono equation \cite{NazSk}.
By multiplying (\ref{Haminf}) by $\beta$, we obtain:
\begin{align}\label{Haminf2}\notag
\beta \H=\beta\oint_{\xi}\frac{d\xi}{\xi}\phi_a^-(\xi)(\phi_a^+(\xi))^2+\beta\oint_{\xi}\frac{d\xi}{\xi}\phi_a^-(\xi)(\phi_a^+(\xi))^\prime+\beta^2\sum_{a,b}\oint_{\eta}\frac{d\eta}{\eta}\phi_a^-(\eta)\phi_b^-(\eta)\phi_a^+(\eta)
\\
+\beta^2\sum_{a>b}\oint_{\eta,\xi}\frac{d\xi d\eta}{\xi\eta}\phi_a^-(\eta)\phi_b^-(\xi)\sum_k k\left(\frac{\xi^k}{\eta^k}+\frac{\eta^k}{\xi^k}\right) V_a^{-1}(\eta)V_b(\eta)V_b^{-1}(\xi)V_a(\xi),
\end{align}
where for convenience we denote $V_a(x)V_b^{-1}(x)$ by $V_{ab}(x)$. 

Introducing the classical variables $\alpha_{n,a},\ n\in \Z$
\begin{equation*}
 \alpha_{n,a}=\begin{cases}
\beta p_{-n,a}, & n\leqslant 0 ;\\
n\frac{\partial}{\partial p_{n,a}}, & n>0,
 \end{cases}
\end{equation*}
with Poisson bracket relations
\begin{equation}\label{pb1}
\{\alpha_{n,a},\alpha_{m,b}\}=n\delta_{a,b}\delta_{n+m,0},
\end{equation}
combine them into generating functions 
$$
\varphi_a^-(\xi)=\sum_{n=0}^{\infty}\frac{\alpha_{-n,a}}{\xi^n},\ \varphi_a^+(\xi)=\sum_{n=1}^{\infty}\alpha_{n,a}\xi^n,
\qquad\text{and}\ \  \varphi_a(x)=\varphi_a^+(x)+\varphi_a^-(x).
$$
Then 
$$
\{\varphi_a^-(x),\varphi_b^+(y)\}=\delta_{a,b}\frac{y}{(1-\frac{y}{x})^2},\qquad 
\{\varphi_a^\pm(x),\varphi_b^\pm(y)\}=0.
$$
or 
$$
\{\varphi_a(x),\varphi_b(y)\}=\delta'\left(x/y\right)\delta_{a,b}.
$$
The classical counterparts  $\mathcal{V}_a(\xi)$   of the vertex operatosr ${V}_a(\xi)$ satisfy the relations
$$
\xi\frac{d \log\mathcal{V}_a(\xi)}{d\xi}=\varphi_a^+(\xi),
$$ and 
\begin{equation}\label{brak}
\{\varphi_a^-(x),\mathcal{V}_b(y)\}=-\delta_{a,b}\frac{\mathcal{V}_b(y)}{1-\frac{y}{x}}, \qquad 
\{\varphi_a^+(x),\mathcal{V}_b(y)\}=0.
\end{equation}
As before, we use the notation $\mathcal{V}_{ab}(\xi)$ for $\mathcal{V}_{a}(\xi)\mathcal{V}_{b}^{-1}(\xi)$. Set 
\begin{align} \label{Hclas} \notag
\mathcal{H}=\oint_{\xi}\frac{d\xi}{\xi}\varphi_a^-(\xi)(\varphi_a^+(\xi))^2+\oint_{\xi}\frac{d\xi}{\xi}\varphi_a^-(\xi)(\varphi_a^+(\xi))^\prime+\sum_{a,b}\oint_{\eta}\frac{d\eta}{\eta}\varphi_a^-(\eta)\varphi_b^-(\eta)\varphi_a^+(\eta)
\\ 
+\sum_{a>b}\oint_{\eta,\xi}\frac{d\xi d\eta}{\xi\eta}\varphi_a^-(\eta)\varphi_b^-(\xi)\sum_k k\left(\frac{\xi^k}{\eta^k}+\frac{\eta^k}{\xi^k}\right) \mathcal{V}_{ab}^{-1}(\eta)\mathcal{V}_{ab}(\xi).
\end{align}

The operator $\mathcal{H}$ is the classical limit of the Hamiltonian (\ref{Haminf2}) $(\beta\rightarrow 0)$. The rule between the quantum commutator and Poisson bracket is 
$
\beta^{-1}[\ ,\ ]\rightarrow \{\ ,\ \}.
$

\begin{Prop}\label{prop6}\footnote{For a formal series $f(z)=\sum_{n\in\ZZ}f_nz^n$ we denote by $f^+(z)$ the series $f^+(z)=\sum_{n\geq 1}f_nz^n=\oint\frac{zf(\xi)d\xi}{(1-z/\xi)\xi}$ and by
$f^-(z)$ the series
$f^-(z)=\sum_{n\leq 0}f_nz^n=\oint\frac{f(\xi)d\xi}{(1-\xi/z)\xi}$.}
 The equations of motion determined by the Hamiltonian $\mathcal{H}$ are:
\begin{align}\label{em+}
\notag
\{\varphi_a^+(x),\mathcal{H}\}=x\frac{\partial}{\partial x}(\varphi_a^+(x))^2+\left(x\frac{\partial}{\partial x}\right)^2\left(\varphi_a^+(x)\right)+\sum_{b}x\frac{\partial}{\partial x}\left(\varphi_b^-(x)\varphi_a^+(x)\right)^++
\\ \notag
+\sum_{b}x\frac{\partial}{\partial x}\left(\varphi_b^-(x)\varphi_b^+(x)\right)^+
+\sum_{b\neq a}x\frac{\partial}{\partial x}
\left(\mathcal{V}_{ab}^{-1}(x)x\frac{\partial}{\partial x}
\left(\left(\varphi_b^-(x)\mathcal{V}_{ab}(x)\right)^{+}-
\left(\varphi_b^-(x)\mathcal{V}_{ab}(x)\right)^{-}\right)\right)^+=
\\ \notag
x\frac{\partial}{\partial x}(\varphi_a^+(x))^2+\left(x\frac{\partial}{\partial x}\right)^2\left(\varphi_a^+(x)
\right)+2\sum_{b}x\frac{\partial}{\partial x}\left(\varphi_b^-(x)\varphi_b^+(x)\right)^+
\\
+2\sum_{b\neq a}x\frac{\partial}{\partial x}\left(\mathcal{V}_{ab}^{-1}(x)x\frac{\partial}{\partial x}
\left(\varphi_b^-(x)\mathcal{V}_{ab}(x)\right)^{+}\right),
\end{align}
\begin{align}\label{em-}
\notag
\{\varphi_a^-(x),\mathcal{H}\}=2x\frac{\partial}{\partial x}\left(\varphi_a^-(x)\varphi_a^+(x)\right)^--\left(x\frac{\partial}{\partial x}\right)^2\left(\varphi_a^-(x)\right)+\sum_{b}x\frac{\partial}{\partial x}\left(\varphi_b^-(x)\varphi_a^-(x)\right)
\\ \notag
+\sum_{b\neq a}\left(\varphi_a^-(x)\mathcal{V}_{ab}^{-1}(x)x\frac{\partial}{\partial x}\left(\left(\varphi_b^-(x)\mathcal{V}_{ab}(x)\right)^{+}-\left(\varphi_b^-(x)\mathcal{V}_{ab}(x)\right)^{-}\right)\right)^-
\\
-\sum_{b\neq a}\left(\varphi_b^-(x)\mathcal{V}_{ab}(x)x\frac{\partial}{\partial x}\left(\left(\varphi_a^-(x)\mathcal{V}_{ab}^{-1}(x)\right)^{+}-\left(\varphi_a^-(x)\mathcal{V}_{ab}^{-1}(x)\right)^{-}\right)\right)^-
\end{align}
\end{Prop}
{\bf Remark} Unlike the Yangian generators \rf{tab0}, \rf{tab1}, the Hamiltonian \rf{Haminf} does not contain dual zero modes $\frac{\partial}{\partial p_{0a}}$. The same holds for the classical limit, where we can freely use the operators $\tilde{\mathcal{V}}_{a}(\xi)= \exp\sum_{n\geq 1}\frac{a_n}{n}\xi^n$ instead of  ${\mathcal{V}}_{a}(\xi)$. The Hamiltonian and the equations of motion do not change, while the brackets \rf{brak} turn into
$$
  \{\varphi_a^-(x),\tilde{\mathcal{V}}_b(y)\}=
  -\delta_{a,b}\frac{y/x\tilde{\mathcal{V}}_b(y)}{1-\frac{y}{x}}, \qquad 
  \{\varphi_a^+(x),\tilde{\mathcal{V}}_b(y)\}=0.
 $$

The quantum system is integrable: it has an infinite number of integrals of motion that can be obtained 
from the 
$q$-determinant of the Yangian generator function $T^{ab}(u)$. It is natural to assume that the classical system is integrable as well. In particular, it should admit a Lax pair presentation. 
Consider the operators $L$ and $M$:
\begin{equation}
\begin{array}{l}
\displaystyle Lf=z\frac{\partial }{\partial z}f(z)+\sum_a \mathcal{V}_{a}(z)\left(\varphi_a^-(z)\mathcal{V}_{a}^{-1}(z)f(z)\right)^+,\\
\displaystyle Mf=\left(z\frac{\partial}{\partial z}\right)^2\!\!f(z)+2\sum_b\left(\varphi_b^+(z)\varphi_b^-(z)\right)^+\!f(z)+2\sum_b \mathcal{V}_{b}(z)z\frac{\partial }{\partial z}\left(\varphi_b^-(z)\mathcal{V}_{b}^{-1}(z)f(z)\right)^+.
\end{array} 
\label{LM}
\end{equation}
They act on the space of analytic functions
$$f(z)=f_0+f_1 z+f_2 z^2+\dots,$$ 
where coefficients $f_i$ are polysymmetrical functions $f_i(p_{n,a})$.
\begin{Prop}
The operators $L$ and $M$ (\ref{LM})  represent a Lax pair of the classical system (\ref{Hclas}):
$$
\frac{dL}{dt}=[M,L].
$$
\end{Prop}
\setcounter{section}{0}
\renewcommand{\thesection}{}
\renewcommand{\thesubsection}{\Alph{section}.\arabic{subsection}}
\section{Appendix}
\subsection{Rational Calogero system}
In this section we present the analogous formulas for the rational Calogero model in the limit of infinitely many particles. We use all notations as before. The Dunkl operators for the rational system are defined as:
\begin{equation*}
D_{i}=\frac{\partial}{\partial x_i}+\beta\sum_{j\neq i,\sigma(i)=\sigma(j)}\frac{1}{x_i-x_j}\left( 1 - K_{ij}\right)
+\beta\sum_{j,\sigma(i)\neq\sigma(j)}\frac{1}{x_i-x_j}\left( 1 - K_{ij}\right).
\end{equation*}
We use a similar construction for the rational case and here present the analog of formulas (\ref{Dinf}) and (\ref{eabinf}):
\begin{equation}
D\left(F_a(z)\right)=\frac{\partial}{\partial z}F_a(z,p)+\beta\sum_b \oint \frac{d\xi}{\xi^2}\frac{\phi_b^-(\xi)}{1-\frac{z}{\xi}}V_b(z)V_b^{-1}(\xi)F_a(\xi),
\end{equation}
\begin{equation}
\E^{ab}\left(F_c(z)\right)=\delta_{bc}\oint_{\xi}\frac{d\xi}{\xi}\phi_a^-(\xi)V_a^{-1}(\xi)F_c(\xi).
\end{equation}
The answer for the second Hamiltonian is 
$$
\H_2=\sum_{k,n\geqslant 1,a}k n p_{n+k-2,a}\frac{\partial}{\partial p_{k,a}}\frac{\partial}{\partial p_{n,a}}+
(1-\beta)\sum_{n>1,a}n(n-1)p_{n-2,a}\frac{\partial}{\partial p_{n,a}}+$$
$$\beta\sum_{a,b}\sum_{n,k\geqslant 0}(k+n+2)p_{k,a} p_{n,b}\frac{\partial}{\partial p_{k+n+2}}+
$$
$$
+\beta\sum_{a>b} \oint\frac{d\xi d\eta}{\xi^2\eta^2}\phi_a^-(\eta)\phi_b^-(\xi)\left(n\frac{\xi^n}{\eta^n}+n\frac{\eta^n}{\xi^n}\right)V_b(\eta)V_b^{-1}(\xi)V_a(\xi)V_a^{-1}(\eta).
$$

\subsection{Calculations of Hamiltonians}
In this section the proofs for formulas (\ref{h11}),(\ref{Haminf}) are presented.
$$
\H_{1}=\sum_i D_i +\frac{\beta}{2}\sum_{a,b}\mathbb{E}^{ab}\mathbb{E}^{ba}-\frac{\beta}{2} s p_0=
$$
$$
=\sum_a\oint\frac{d\xi}{\xi}\phi_a^-(\xi)V_a^{-1}(\xi)\left(\xi\frac{\partial}{\partial \xi}\right)V_a(\xi)+\beta\sum_{a,b} \oint\frac{d\xi d\eta}{\xi\eta}\phi_a^-(\eta)V_a^{-1}(\eta)\frac{\phi_b^-(\xi)\frac{\eta}{\xi}}{1-\frac{\eta}{\xi}}V_b(\eta)V_b^{-1}(\xi)V_a(\xi)
$$
$$
+\frac{\beta}{2}\sum_{a,b} \oint\frac{d\xi d\eta}{\xi\eta}\phi_a^-(\eta)V_a^{-1}(\eta)V_b(\eta)\phi_b^-(\xi)V_b^{-1}(\xi)V_a(\xi)-\frac{\beta}{2} s p_0
$$
We split the second sum into two parts $a\neq b$ and $a=b$, swap $V_b(\eta)$ and $\phi_b^-(\xi)$ in the third item using commutator relations $[V_b(\eta),\phi_b^-(\xi)]=\frac{1}{1-\frac{\eta}{\xi}}V_b(\eta)$.
\begin{equation*}\begin{split}
\H_1=&\sum_a\oint\frac{d\xi}{\xi}\phi_a^-(\xi)\phi_a^+(\xi)+\beta\sum_{a\neq b} \oint\frac{d\xi d\eta}{\xi\eta}\phi_a^-(\eta)\phi_b^-(\xi)\frac{\frac{\eta}{\xi}}{1-\frac{\eta}{\xi}}V_a^{-1}(\eta)V_b(\eta)V_b^{-1}(\xi)V_a(\xi)+
\\
+&\beta\sum_{a}\oint\frac{d\xi d\eta}{\xi\eta}\phi_a^-(\eta)V_a^{-1}(\eta)\phi_a^-(\xi)
\frac{\frac{\eta}{\xi}}{1-\frac{\eta}{\xi}}V_a(\eta)
\\+&\frac{\beta}{2}\sum_{a,b}\oint\frac{d\xi d\eta}{\xi\eta}\phi_a^-(\eta)V_a^{-1}(\eta)\phi_b^-(\xi)V_b(\eta)V_b^{-1}(\xi)V_a(\xi)
\\
+&\frac{\beta}{2}\sum_{a,b}\oint\frac{d\xi d\eta}{\xi\eta}\phi_a^-(\eta)V_a^{-1}(\eta)\frac{1}{1-\frac{\eta}{\xi}}V_b(\eta)V_b^{-1}(\xi)V_a(\xi)-\frac{\beta}{2} s p_0
\end{split}\end{equation*}
The third item vanishes as $\xi$ enters only in negative powers. The last integral can be simplified
$$
\oint\frac{d\xi d\eta}{\xi\eta}\phi_a^-(\eta)V_a^{-1}(\eta)\frac{1}{1-\frac{\eta}{\xi}}V_b(\eta)V_b^{-1}(\xi)V_a(\xi)=\oint\frac{d\xi d\eta}{\xi\eta}\phi_a^-(\eta)V_a^{-1}(\eta)V_b(\eta)V_b^{-1}(\eta)V_a(\eta)=p_{0,a}
$$
Thus it cancels with $-\frac{\beta}{2} s p_0$. We halve the second integral and change the indices of summing $a,b$ and variables $\xi,\eta$ in the second one. Thus we obtain
$$
\H_1=\sum_a\oint\frac{d\xi}{\xi}\phi_a^-(\xi)\phi_a^+(\xi)+\frac{\beta}{2}\sum_{a\neq b} \oint\frac{d\xi d\eta}{\xi\eta}\phi_a^-(\eta)\phi_b^-(\xi)\frac{\frac{\eta}{\xi}}{1-\frac{\eta}{\xi}}V_a^{-1}(\eta)V_b(\eta)V_b^{-1}(\xi)V_a(\xi)+
$$
$$
+\frac{\beta}{2}\sum_{a\neq b} \oint\frac{d\xi d\eta}{\xi\eta}\phi_b^-(\xi)\phi_a^-(\eta)\frac{\frac{\xi}{\eta}}{1-\frac{\xi}{\eta}}V_b^{-1}(\xi)V_a(\xi)V_a^{-1}(\eta)V_b(\eta)
$$
$$
+\frac{\beta}{2}\sum_{a\neq b}\oint\frac{d\xi d\eta}{\xi\eta}\phi_a^-(\eta)V_a^{-1}(\eta)\phi_b^-(\xi)V_b(\eta)V_b^{-1}(\xi)V_a(\xi)+\frac{\beta}{2}\sum_{a}\oint\frac{d\xi d\eta}{\xi\eta}\phi_a^-(\eta)V_a^{-1}(\eta)\phi_a^-(\xi)V_a(\eta)
$$
The second, third and forth integrals are combined into an integral with delta-function $\delta\left(\frac{\xi}{\eta}\right)=\sum_{n\in\mathbb{Z}}\frac{\xi^n}{\eta^n}$, which is expressed by the following formula
$$
\oint\frac{d\xi d\eta}{\xi\eta}\phi_a^-(\eta)\phi_b^-(\xi)\delta\left(\frac{\xi}{\eta}\right)V_a^{-1}(\eta)V_b(\eta)V_b^{-1}(\xi)V_a(\xi)=\oint\frac{d\xi d\eta}{\xi\eta}\phi_a^-(\eta)\phi_b^-(\xi)=p_{0,a}p_{0,b},
$$
while the last integral equals $p_{0,a}^2-p_{0,a}$. The final answer is the following
$$
\H_1=\sum_a\oint\frac{d\xi}{\xi}\phi_a^-(\xi)\phi_a^+(\xi)+\frac{\beta}{2}\left(p_{0}^2-p_{0}\right), 
$$
where $p_0=\sum_a p_{0,a}$.
Now we calculate the second Hamiltonian:
\begin{equation}\label{hcon}\begin{split}
\H_{2}=&\sum_i D_i^2+\beta\sum_{i,j,a,b}E^{ab}_iE^{ba}_jD_j-\beta s\sum_i D_i
+\frac{\beta^2}{3}\sum_{a,b,c}\mathbb{E}^{ab}\mathbb{E}^{bc}\mathbb{E}^{ca}\\
-&\frac{2s\beta^2}{3}\sum_{a,b}\mathbb{E}^{ab}\mathbb{E}^{ba}+\beta^2\frac{2s^2+p_0-1}{6}p_0,
\end{split}\end{equation}
In this expression we separately calculate the operator coefficients at $\beta^0,\beta, \beta^2$:
$$
\H=\H_{2}-\beta\H_1=U_0+\beta U_1+\beta^2 U_2
$$
$$
U_0=\sum_a\oint\frac{d\xi}{\xi}\phi_a^-(\xi)V_a^{-1}(\xi)\left(\xi\frac{\partial}{\partial \xi}\right)^2V_a(\xi)=\sum_a\oint\frac{d\xi}{\xi}\phi_a^-(\xi)(\phi_a^+(\xi))^2+\sum_a\oint\frac{d\xi}{\xi}\phi_a^-(\xi)(\phi_a^+(\xi))^\prime
$$
The first three items in (\ref{hcon}) contribute to $U_1$:
$$
U_1=\sum_{a,b}\oint\frac{d\xi d\eta}{\xi\eta}\phi_a^-(\eta)V_a^{-1}(\eta)\phi_b^-(\xi)\left(\frac{\phi_b^+(\eta)\frac{\eta}{\xi}}{1-\frac{\eta}{\xi}}+\frac{\frac{\eta}{\xi}}{(1-\frac{\eta}{\xi})^2} \right)V_b(\eta)V_b^{-1}(\xi)V_a(\xi)+
$$
$$
+\sum_{a,b}\oint\frac{d\xi d\eta}{\xi\eta}\phi_a^-(\eta)V_a^{-1}(\eta)\phi_b^-(\xi)\frac{\frac{\eta}{\xi}}{1-\frac{\eta}{\xi}}V_b(\eta)V_b^{-1}(\xi)\phi_a^+(\xi)V_a(\xi)+
$$
$$
+\sum_{a,b}\oint\frac{d\xi d\eta}{\xi\eta}\phi_a^-(\eta)V_a^{-1}(\eta)V_b(\eta)\phi_b^-(\xi)V_b^{-1}(\xi)\phi_a^+(\xi)V_a(\xi)-(s+1)\sum_a \oint\frac{d\xi}{\xi}\phi_a^-(\xi)\phi_a^+(\xi)=
$$
In the first item we can take the sum $a\neq b$ as in case $a=b$ it vanishes. In the third integral we use the commutator relations $[V_b(\eta),\phi_b^-(\xi)]=\frac{1}{1-\frac{\eta}{\xi}}V_b(\eta)$ and add it to the second item.
$$
U_1=\sum_{a\neq b}\oint\frac{d\xi d\eta}{\xi\eta}\phi_a^-(\eta)V_a^{-1}(\eta)\phi_b^-(\xi)\left(\frac{\phi_b^+(\eta)\frac{\eta}{\xi}}{1-\frac{\eta}{\xi}}+\frac{\frac{\eta}{\xi}}{(1-\frac{\eta}{\xi})^2} \right)V_b(\eta)V_b^{-1}(\xi)V_a(\xi)+
$$
$$
+\sum_{a,b}\oint\frac{d\xi d\eta}{\xi\eta}\phi_a^-(\eta)V_a^{-1}(\eta)\phi_b^-(\xi)\frac{1}{1-\frac{\eta}{\xi}}V_b(\eta)V_b^{-1}(\xi)\phi_a^+(\xi)V_a(\xi)+
$$
$$
+\sum_{a,b}\oint\frac{d\xi d\eta}{\xi\eta}\phi_a^-(\eta)V_a^{-1}(\eta)\frac{1}{1-\frac{\eta}{\xi}}V_b(\eta)V_b^{-1}(\xi)\phi_a^+(\xi)V_a(\xi)-(s+1)\sum_a \oint\frac{d\xi}{\xi}\phi_a^-(\xi)\phi_a^+(\xi)=
$$
In the second item we change the indices of summing $a,b$ and variables $\xi,\eta$ and as in $H_1$ we combine it with the first integral and obtain the delta-function. The third integral can be simplified:
$$
\oint\frac{d\eta}{\eta}\phi_a^-(\eta)V_a^{-1}(\eta)V_b(\eta)V_b^{-1}(\eta)\phi_a^+(\eta)V_a(\eta)=\oint\frac{d\eta}{\eta}\phi_a^-(\eta)\phi_a^+(\eta)
$$
Thus we obtain
$$
U_1=\sum_{a\neq b}\oint\frac{d\xi d\eta}{\xi\eta}\phi_a^-(\eta)V_a^{-1}(\eta)\phi_b^-(\xi)\delta\left(\frac{\xi}{\eta}\right)V_b(\eta)V_b^{-1}(\xi)\phi_a^+(\xi)V_a(\xi)
$$
$$
+\sum_{a\neq b}\oint\frac{d\xi d\eta}{\xi\eta}\phi_a^-(\eta)V_a^{-1}(\eta)\phi_b^-(\xi)\frac{\frac{\eta}{\xi}}{(1-\frac{\eta}{\xi})^2}V_b(\eta)V_b^{-1}(\xi)V_a(\xi)-\sum_a \oint\frac{d\xi}{\xi}\phi_a^-(\xi)\phi_a^+(\xi)
$$
$$
+\sum_{a}\oint\frac{d\xi d\eta}{\xi\eta}\phi_a^-(\eta)V_a^{-1}(\eta)\phi_a^-(\xi)\frac{1}{1-\frac{\eta}{\xi}}V_a(\eta)\phi_a^+(\xi)
$$
Using commutator relations $[V_a^{-1}(\eta),\phi_a^-(\xi)]=-\frac{1}{1-\frac{\eta}{\xi}}V_a^{-1}(\eta)$ in the last item we obtain 
$$
U_1=\sum_{a\neq b}\oint\frac{d\xi}{\xi}\phi_a^-(\xi)\phi_b^-(\xi)\phi_a^+(\xi)-\sum_a \oint\frac{d\xi}{\xi}\phi_a^-(\xi)\phi_a^+(\xi)
$$
$$
+\sum_{a>b}\oint_{\eta,\xi}\frac{d\xi d\eta}{\xi\eta}\phi_a^-(\eta)\phi_b^-(\xi)\sum_k k\left(\frac{\xi^k}{\eta^k}+\frac{\eta^k}{\xi^k}\right) V_a^{-1}(\eta)V_b(\eta)V_b^{-1}(\xi)V_a(\xi)
$$
$$
+\sum_{a}\oint\frac{d\xi}{\xi}\phi_a^-(\xi)\phi_a^-(\xi)\phi_a^+(\xi)-\sum_{a}\oint\frac{d\xi d\eta}{\xi\eta}\phi_a^-(\eta)\frac{1}{(1-\frac{\eta}{\xi})^2}\phi_a^+(\xi)
$$
$$
U_1=\sum_{a, b}\oint\frac{d\xi}{\xi}\phi_a^-(\xi)\phi_b^-(\xi)\phi_a^+(\xi)-
\sum_{a}\oint\frac{d\xi}{\xi\eta}\phi_a^-(\xi)(\phi_a^+(\xi))'-\sum_a \oint\frac{d\xi}{\xi}\phi_a^-(\xi)\phi_a^+(\xi)+
$$
$$
\sum_{a>b}\oint_{\eta,\xi}\frac{d\xi d\eta}{\xi\eta}\phi_a^-(\eta)\phi_b^-(\xi)\sum_k k\left(\frac{\xi^k}{\eta^k}+
\frac{\eta^k}{\xi^k}\right) V_a^{-1}(\eta)V_b(\eta)V_b^{-1}(\xi)V_a(\xi)
$$
$$
U_2=\sum_{a,b,c} \oint\frac{d\xi d\eta d\gamma}{\xi\eta\gamma}\frac{\frac{\gamma}{\eta}}{1-\frac{\gamma}{\eta}}
\frac{\frac{\eta}{\xi}}{1-\frac{\eta}{\xi}}\phi_a^-(\gamma)V_a^{-1}(\gamma)\phi_c^-(\eta)
V_c(\gamma)V_c^{-1}(\eta)\phi_b^-(\xi)V_b(\eta)V_b^{-1}(\xi)V_a(\xi)
$$
$$
+\sum_{a,b,c} \oint\frac{d\xi d\eta d\gamma}{\xi\eta\gamma}\frac{\frac{\eta}{\xi}}{1-\frac{\eta}{\xi}}\phi_a^-(\gamma)V_a^{-1}(\gamma)V_c(\gamma)\phi_c^-(\eta)V_c^{-1}(\eta)\phi_b^-(\xi)V_b(\eta)V_b^{-1}(\xi)V_a(\xi)
$$
$$
-s\sum_{a,b} \oint\frac{d\xi d\eta}{\xi\eta}\frac{\frac{\eta}{\xi}}{1-\frac{\eta}{\xi}}\phi_a^-(\eta)V_a^{-1}(\eta)\phi_b^-(\xi)V_b(\eta)V_b^{-1}(\xi)V_a(\xi)
$$
$$
+\frac{1}{3}\sum_{a,b,c} \oint\frac{d\xi d\eta d\gamma}{\xi\eta\gamma}\phi_a^-(\gamma)V_a^{-1}(\gamma)V_c(\gamma)\phi_c^-(\eta)V_c^{-1}(\eta)V_b(\eta)\phi_b^-(\xi)V_b^{-1}(\xi)V_a(\xi)
$$
$$
-\frac{2}{3}\sum_{a,b} \oint\frac{d\xi d\eta}{\xi\eta}\phi_a^-(\eta)V_a^{-1}(\eta)V_b(\eta)\phi_b^-(\xi)V_b^{-1}(\xi)V_a(\xi)+\frac{s^2-p_0+1}{3}p_0
$$
\subsection{Calculations for classical systems}
 Here we provide a calculation for the Poisson brackets in the classical limit.
$$
\{\varphi_a(x),\varphi_b(y)\}=\delta'\left(x/y\right)\delta_{a,b}.
$$
Calculate Poisson bracket with the Hamiltonian $\{\phi_a^-(x),\mathcal{H}\}$, $\{\phi_a^+(x),\mathcal{H}\}$:
$$
\{\varphi_a^-(x),\oint_{\xi}\frac{d\xi}{\xi}\varphi_a^-(\xi)(\varphi_a^+(\xi))^2\}=-2\oint_{\xi}\frac{d\xi}{\xi}\varphi_a^-(\xi)\varphi_a^+(\xi)\sum_{n>0}n\frac{\xi^n}{x^n}
$$
$$
\{\varphi_a^+(x),\oint_{\xi}\frac{d\xi}{\xi}\varphi_a^-(\xi)(\varphi_a^+(\xi))^2\}=\oint_{\xi}\frac{d\xi}{\xi}(\varphi_a^+(\xi))^2\sum_{n>0}n\frac{x^n}{\xi^n}=x\frac{d}{dx}(\varphi_a^+(x))^2
$$
$$
\{\varphi_a^-(x),\oint_{\xi}\frac{d\xi}{\xi}(\varphi_a^-(\xi))^2\varphi_a^+(\xi)\}=-\oint_{\xi}\frac{d\xi}{\xi}(\varphi_a^-(\xi))^2\sum_{n>0}n\frac{\xi^n}{x^n}=x\frac{d}{dx}(\varphi_a^-(x))^2
$$
$$
\{\varphi_a^+(x),\oint_{\xi}\frac{d\xi}{\xi}(\varphi_a^-(\xi))^2\varphi_a^+(\xi)\}=
2\oint_{\xi}\frac{d\xi}{\xi}\varphi_a^-(\xi)\varphi_a^+(\xi)\sum_{n>0}n\frac{x^n}{\xi^n}
$$
From these four formulas we obtain 
$$
\{\varphi_a(x),\oint_{\xi}\frac{d\xi}{\xi}\left((\varphi_a^-(\xi))^2\varphi_a^+(\xi)+\varphi_a^-(\xi)(\varphi_a^+(\xi))^2\right)\}=x\frac{d}{dx}(\varphi_a(x))^2.
$$
Summing the following two equations
$$
\{\varphi_a^-(x),\oint_{\xi}\frac{d\xi}{\xi}\varphi_a^-(\xi)(\varphi_a^+(\xi))^\prime\}=-\oint_{\xi}\frac{d\xi}{\xi}\varphi_a^-(\xi)\xi\frac{d}{d\xi}\left(\sum_{n>0}n\frac{\xi^n}{x^n}\right)=-\left(x\frac{d}{dx}\right)^2\varphi_a^-(x)
$$
$$
\{\varphi_a^+(x),\oint_{\xi}\frac{d\xi}{\xi}\varphi_a^-(\xi)(\varphi_a^+(\xi))^\prime\}=\oint_{\xi}\frac{d\xi}{\xi}(\varphi_a^+(\xi))'\left(\sum_{n>0}n\frac{x^n}{\xi^n}\right)=\left(x\frac{d}{dx}\right)^2\varphi_a^+(x)
$$
we obtain
$$
\{\varphi_a(x),\oint_{\xi}\frac{d\xi}{\xi}\varphi_a^-(\xi)(\varphi_a^+(\xi))^\prime\}=\left(x\frac{d}{dx}\right)^2\left(\varphi_a^+(x)-\varphi_a^-(x)\right).
$$
The next items are the following
$$
\{\varphi_a^-(x),\oint_{\eta}\frac{d\eta}{\eta}\varphi_a^-(\eta)\varphi_b^-(\eta)\varphi_a^+(\eta)\}=-\oint_{\eta}\frac{d\eta}{\eta}\varphi_a^-(\eta)\varphi_b^-(\eta)\sum_{n>0}n\frac{\eta^n}{x^n}=x\frac{d}{dx}\left(\varphi_a^-(x)\varphi_b^-(x)\right)
$$
$$
\{\varphi_a^+(x),\oint_{\eta}\frac{d\eta}{\eta}\varphi_a^-(\eta)\varphi_b^-(\eta)\varphi_a^+(\eta)\}=\oint_{\eta}\frac{d\eta}{\eta}\varphi_a^+(\eta)\varphi_b^-(\eta)\sum_{n>0}n\frac{x^n}{\eta^n}=x\frac{d}{dx}\left(\varphi_a^+(x)\varphi_b^-(x)\right)^+
$$
$$
\{\varphi_a^+(x),\oint_{\eta}\frac{d\eta}{\eta}\varphi_b^-(\eta)\varphi_a^-(\eta)\varphi_b^+(\eta)\}=x\frac{d}{dx}\left(\varphi_b^+(x)\varphi_b^-(x)\right)^+
$$
Poisson brackets with the the double integral are:
$$
\{\varphi_a^-(x),\oint_{\eta,\xi}\frac{d\xi d\eta}{\xi\eta}\varphi_a^-(\eta)\varphi_b^-(\xi)
\sum_k k\left(\frac{\xi^k}{\eta^k}+\frac{\eta^k}{\xi^k}\right) 
\mathcal{V}_a^{-1}(\eta)\mathcal{V}_b(\eta)\mathcal{V}_b^{-1}(\xi)\mathcal{V}_a(\xi)\}
$$
$$
=\oint_{\eta,\xi}\frac{d\xi d\eta}{\xi\eta}\left(\frac{1}{1-\frac{\eta}{x}}-\frac{1}{1-\frac{\xi}{x}}\right)
\varphi_a^-(\eta)\varphi_b^-(\xi)\sum_k k\left(\frac{\xi^k}{\eta^k}+\frac{\eta^k}{\xi^k}\right) 
\mathcal{V}_a^{-1}(\eta)\mathcal{V}_b(\eta)\mathcal{V}_b^{-1}(\xi)\mathcal{V}_a(\xi)
$$
$$
=\oint_{\eta}\frac{d\eta}{\eta}\frac{1}{1-\frac{\eta}{x}}\mathcal{V}_a^{-1}(\eta)
\mathcal{V}_b(\eta)\varphi_a^-(\eta)\left(\eta\frac{d}{d\eta}\left(\varphi_b^-(\eta)
\mathcal{V}_b^{-1}(\eta)\mathcal{V}_a(\eta)\right)^{\eta+}-\eta\frac{d}{d\eta}
\left(\varphi_b^-(\eta)\mathcal{V}_b^{-1}(\eta)\mathcal{V}_a(\eta)\right)^{\eta-}\right)
$$
$$
-\oint_{\xi}\frac{d\xi}{\xi}\frac{1}{1-\frac{\xi}{x}}\varphi_b^-(\xi)\mathcal{V}_b^{-1}(\xi)\mathcal{V}_a(\xi)
\left(\xi\frac{\partial}{\partial\xi}\left(\varphi_a^-(\xi)\mathcal{V}_a^{-1}(\xi)\mathcal{V}_b(\xi)\right)^{\xi+}-
\xi\frac{\partial}{\partial\xi}\left(\varphi_a^-(\xi)\mathcal{V}_a^{-1}(\xi)\mathcal{V}_b(\xi)\right)^{\xi-}\right)
$$
$$
=\left(\varphi_a^-(x)\mathcal{V}_a^{-1}(x)\mathcal{V}_b(x)x\frac{\partial}{\partial x}\left(\left(\varphi_b^-(x)
\mathcal{V}_b^{-1}(x)\mathcal{V}_a(x)\right)^{+}-\left(\varphi_b^-(x)\mathcal{V}_b^{-1}(x)\mathcal{V}_a(x)\right)^{-}\right)\right)^-
$$
$$
-\left(\varphi_b^-(x)\mathcal{V}_b^{-1}(x)\mathcal{V}_a(x)x\frac{\partial}{\partial x}\left(\left(\varphi_a^-(x)\mathcal{V}_a^{-1}(x)
\mathcal{V}_b(x)\right)^{+}-\left(\varphi_a^-(x)\mathcal{V}_a^{-1}(x)\mathcal{V}_b(x)\right)^{-}\right)\right)^-.
$$
$$
\{\varphi_a^+(x),\oint_{\eta,\xi}\frac{d\xi d\eta}{\xi\eta}\varphi_a^-(\eta)\varphi_b^-(\xi)
\sum_k k\left(\frac{\xi^k}{\eta^k}+\frac{\eta^k}{\xi^k}\right) \mathcal{V}_a^{-1}(\eta)\mathcal{V}_b(\eta)
\mathcal{V}_b^{-1}(\xi)\mathcal{V}_a(\xi)\}
$$
$$
=\oint_{\eta,\xi}\frac{d\xi d\eta}{\xi\eta}\sum_{n>0}n\frac{x^n}{\eta^n}\varphi_b^-(\xi)\sum_k k\left(\frac{\xi^k}{\eta^k}+
\frac{\eta^k}{\xi^k}\right) \mathcal{V}_a^{-1}(\eta)\mathcal{V}_b(\eta)\mathcal{V}_b^{-1}(\xi)\mathcal{V}_a(\xi)
$$
$$
=\oint_{\eta}\frac{d\eta}{\eta}\sum_{n>0}n\frac{x^n}{\eta^n}\mathcal{V}_a^{-1}(\eta)\mathcal{V}_b(\eta)\left(\eta\frac{d}{d\eta}
\left(\varphi_b^-(\eta)\mathcal{V}_b^{-1}(\eta)\mathcal{V}_a(\eta)\right)^{\eta+}-\eta\frac{d}{d\eta}
\left(\varphi_b^-(\eta)\mathcal{V}_b^{-1}(\eta)\mathcal{V}_a(\eta)\right)^{\eta-}\right)
$$
$$
=x\frac{\partial}{\partial x}\left(\mathcal{V}_a^{-1}(x)\mathcal{V}_b(x)x\frac{\partial}{\partial x}\left(\left(\varphi_b^-(x)\mathcal{V}_b^{-1}(x)\mathcal{V}_a(x)\right)^{+}-\left(\varphi_b^-(x)\mathcal{V}_b^{-1}(x)\mathcal{V}_a(x)\right)^{-}\right)\right)^+.
$$
The final answer is given in proposition \ref{prop6}.
Both sides of the equality 
$
\dfrac{dL}{dt}=[M,L]
$ are equal to
$$
\sum_a \mathcal{V}_a \left(\varphi_a^+\right)^2\left(\varphi_a^-\mathcal{V}_a^{-1}f\right)^+
+\sum_a \mathcal{V}_a z\frac{\partial \varphi_a^+}{\partial z}\left(\varphi_a^-\mathcal{V}_a^{-1}f\right)^+
+2\sum_a \mathcal{V}_a \left(z\frac{\partial}{\partial z}\left(\varphi_a^-\varphi_a^+\right)^-\mathcal{V}_a^{-1}f\right)^+
$$
$$
-\sum_a \mathcal{V}_a \left(\left(z\frac{\partial}{\partial z}\right)^2\left(\varphi_a^-\right)\mathcal{V}_a^{-1}f\right)^+
-\sum_a \mathcal{V}_a \left(\left(\varphi_a^+\right)^2\varphi_a^-\mathcal{V}_a^{-1}f\right)^+
-\sum_a \mathcal{V}_a\left(z\frac{\partial \varphi_a^+}{\partial z}\varphi_a^-\mathcal{V}_a^{-1}f\right)^+
$$
$$
+\sum_a \mathcal{V}_a\left(z\frac{\partial (\varphi_a^-)^2}{\partial z}\mathcal{V}_a^{-1}f\right)^++2\sum_{a,b} 
\mathcal{V}_a (\varphi_b^-\varphi_b^+)_+\left(\varphi_a^-\mathcal{V}_{a}^{-1}f\right)^+-2\sum_{a,b} 
\mathcal{V}_a \left((\varphi_b^-\varphi_b^+)_+\varphi_a^-\mathcal{V}_{a}^{-1}f\right)^+
$$
$$
+2\sum_{a\neq b}\mathcal{V}_{b}\left(z\frac{\partial}{\partial z}(\varphi_b^-
\mathcal{V}_{ab})\left(\varphi_a^-\mathcal{V}_{a}^{-1}f\right)^+\right)^+.
$$
\subsection{  Hamiltonians via $q$-determinant}
Since  $
d_i=D_i+\beta\sum_{j<i}K_{ij}
$, \    
$
H_{1}=\sum_i d_i=\sum_i D_i +\beta\sum_{i,j<i}K_{ij}=\sum_i D_i +\frac{\beta}{2}\sum_{i\neq j}K_{ij}.
$
Here we use the relation $K_{ij}f=P_{ij}f=\sum_{a,b}E^{ab}_iE^{ba}_jf$ on functions $f\in M_N^{S_N}$.
$$
H_{1}=\sum_i D_i +\frac{\beta}{2}\sum_{i\neq j}P_{ij}=\sum_i D_i +\frac{\beta}{2}\sum_{i\neq j,a,b}E^{ab}_iE^{ba}_j=\sum_i D_i +\frac{\beta}{2}\sum_{a,b}\mathbb{E}^{ab}\mathbb{E}^{ba}-\frac{\beta}{2} s p_0.
$$
In the last equality we use $\sum_i E^{ab}_iE^{ba}_i=s p_0$, where $s$ and $p_0=\sum_a p_{0,a}$ is the number of spins and particles, respectively.
$$
H_{2}=\sum_i D_i^2+\beta\sum_{i,j<i}K_{ij}D_i+\beta\sum_{i,j<i}D_iK_{ij}+\beta^2\sum_{i,j<i,k<i}K_{ij}K_{ik}.
$$
Using relations (\ref{1}) in the third item we obtain
$$
H_{2}=\sum_i D_i^2+\beta\sum_{i \neq j}K_{ij}D_i+\beta^2\sum_{i,j<i,k<i}K_{ij}K_{ik}.
$$
The last item can be simplified using commutator relations $E^{ab}_iE^{cd}_j=\delta_{ij}(\delta_{bc}E^{ad}_i-\delta_{ad}E^{bc}_i)$:
$$\sum_{i,j<i,k<i}K_{ij}K_{ik}= \frac{1}{3}\sum_{i\neq j,i\neq k, j\neq i}P_{ij}P_{ik}+\frac{p_0(p_0-1)}{2}=\frac{1}{3}\sum_{i\neq j,i\neq k, j\neq i,a,b,c,d}E^{ab}_iE^{ba}_jE^{cd}_iE^{dc}_k+\frac{p_0(p_0-1)}{2}=$$
$$
=\frac{1}{3}\sum_{i\neq j,i\neq k, j\neq i,a,b,c}E^{ab}_iE^{bc}_jE^{ca}_k+\frac{p_0(p_0-1)}{2}=
$$
$$
=\frac{1}{3}\sum_{i,j,k,a,b,c}E^{ab}_iE^{bc}_jE^{ca}_k-\frac{2s}{3}\sum_{i,j,a,b}E^{ab}_iE^{ba}_j-\frac{1}{3}\sum_{i,j,a,b}E^{aa}_iE^{bb}_j+\frac{s^2 p_0}{3}+\frac{p_0}{3}+\frac{p_0(p_0-1)}{2}.
$$
Thus we obtain
\begin{equation*}\begin{split}
H_{2}=&\sum_i D_i^2+\beta\sum_{i,j,a,b}E^{ab}_iE^{ba}_jD_j-\beta s\sum_i D_i+\frac{\beta^2}{3}\sum_{a,b,c}\mathbb{E}^{ab}\mathbb{E}^{bc}\mathbb{E}^{ca}
\\-&\frac{2s\beta^2}{3}\sum_{a,b}\mathbb{E}^{ab}\mathbb{E}^{ba}+\beta^2\frac{2s^2+p_0-1}{6}p_0.
\end{split}\end{equation*}
Now we show how to obtain the first elements of the $q$-determinant 
$$
 \text{q-det}\, T(u)= \sum_{\sigma\in S_m}(-1)^{sgn(\sigma)}
T^{\sigma(1),1}(u)T^{\sigma(2),2}(u-1) ... T^{\sigma(m),m}(u-m+1)=1+\frac{\Delta_0}{u}+\frac{\Delta_1}{u^2}+\frac{\Delta_2}{u^3}+\dots
$$
where $
T^{ab}(u)=\delta^{ab}+\frac{T^{ab}_0}{u}+\frac{T^{ab}_1}{u^2}+\frac{T^{ab}_2}{u^3}+\dots
$, 
$
\Delta_0=\sum_a T^{aa}_0
$, and 
$$
\Delta_1=-\sum_{a>b} T^{ab}_0  T^{ba}_0 +\sum_{a<b} T^{aa}_0  T^{bb}_0+\sum_a  T^{aa}_1+\sum_{a> 1} (a-1) T^{aa}_0=-\frac{1}{2}\sum_{a,b} T^{ab}_0  T^{ba}_0+\frac{1}{2}\sum_{a,b} T^{aa}_0  T^{bb}_0
$$
$$
-\frac{1}{2}\sum_{a>b}( T^{aa}_0 - T^{bb}_0)+\sum_a  T^{aa}_1+\sum_{a> 1}^s (a-1)T^{aa}_0=\sum_a  T^{aa}_1-\frac{1}{2}\sum_{a,b} T^{ab}_0  T^{ba}_0+\frac{1}{2}(\Delta_0^2+(s-1)\Delta_0),
$$
$$
\Delta_2=\sum_a T^{aa}_2+\sum_{a>1}2(a-1)T^{aa}_1+\sum_{a<b}(T^{aa}_0T^{bb}_1+T^{aa}_1T^{bb}_0)-\sum_{a>b}(T^{ab}_0T^{ba}_1+T^{ab}_1T^{ba}_0)+
$$
$$
\sum_{a<b}(a+b-2)T^{aa}_0T^{bb}_0-\sum_{a>b}(a+b-2)T^{ab}_0T^{ba}_0+\sum_{a>1}(a-1)^2T^{aa}_0+\sum_{a>b>c,\sigma\in S_3}(-1)^{sgn(\sigma)}T^{\sigma(c)c}_0T^{\sigma(b)b}_0T^{\sigma(a)a}_0.
$$
Using the relation $ [T^{ab}_1,  T^{cd}_0]=\delta_{cb}T^{ad}_1-\delta_{ad}T^{cb}_1$ we obtain
$$
\Delta_2=\sum_a T^{aa}_2-\sum_{a,b}T^{ab}_0T^{ba}_1+(s-1)\sum_{a>1}T^{aa}_1+\sum_{a,b}T^{aa}_0T^{bb}_1+
$$
$$
+\sum_{a<b}(a+b-2)T^{aa}_0T^{bb}_0-\sum_{a>b}(a+b-2)T^{ab}_0T^{ba}_0+\sum_{a>1}(a-1)^2T^{aa}_0+\frac{1}{6}\sum_{a,b,c}T^{aa}_0T^{bb}_0T^{cc}_0
$$
$$
+\frac{1}{3}\sum_{a,b,c}T^{ab}_0T^{bc}_0T^{ca}_0-\frac{1}{2}\sum_{a,b,c}T^{cc}_0T^{ab}_0T^{ba}_0+\frac{1}{6}\sum_{a,b}T^{aa}_0T^{bb}_0-\frac{1}{3}\sum_{a,b}T^{ab}_0T^{ba}_0+\frac{2-s}{6}\sum_{a}T^{aa}_0T^{aa}_0
$$
$$
+\frac{1}{3}\sum_{a>b}(3a+3b-1-4s)T^{ab}_0T^{ba}_0-\sum_{b>a}(a+b-1-s)T^{aa}_0T^{bb}_0+\frac{1}{3}\sum_{a>b}-2(s-a)(T^{bb}_0-T^{aa}_0)
$$
$$
=\sum_a T^{aa}_2-\sum_{a,b}T^{ab}_0T^{ba}_1+\frac{1}{3}\sum_{a,b,c}T^{ab}_0T^{bc}_0T^{ca}_0+(s-1)\sum_{a>1}T^{aa}_1+\Delta_0\sum_{a}T^{aa}_1-\Delta_0\frac{1}{2}\sum_{a,b}T^{ab}_0T^{ba}_0
$$
$$
+(s-1)\sum_{a<b}T^{aa}_0T^{bb}_0+\frac{5-4s}{3}\sum_{a>b}T^{ab}_0T^{ba}_0+\frac{1}{6}\sum_{a,b}T^{aa}_0T^{bb}_0-\frac{1}{3}\sum_{a,b}T^{ab}_0T^{ba}_0+\frac{2-s}{6}\sum_{a}T^{aa}_0T^{aa}_0
$$
$$
+\sum_{a>1}(a-1)^2T^{aa}_0+\frac{1}{3}\sum_a (a-s)(s-a-1)T^{aa}_0+\frac{1}{3}\sum_{a}2(s-a)(a-1)T^{aa}_0+\frac{1}{6}\Delta_0^3.
$$
After these calculations we obtain the following formula for $\Delta_2$:
$$
\Delta_2=\sum_a T^{aa}_2-\sum_{a,b}T^{ab}_0T^{ba}_1+\frac{1}{3}\sum_{a,b,c}T^{ab}_0T^{bc}_0T^{ca}_0+\Delta_0\sum_{a}T^{aa}_1-\Delta_0\frac{1}{2}\sum_{a,b}T^{ab}_0T^{ba}_0
$$
$$
(s-1)\sum_{a>1}T^{aa}_1+\frac{3-4s}{6}\sum_{a,b}T^{ab}_0T^{ba}_0+\frac{1}{6}\Delta_0^3+\frac{3s-2}{6}\Delta_0^2+\frac{(s-1)(2s-1)}{6}\Delta_0,
$$
$$
\Delta_2=\sum_a T^{aa}_2-\sum_{a,b}T^{ab}_0T^{ba}_1+\frac{1}{3}\sum_{a,b,c}T^{ab}_0T^{bc}_0T^{ca}_0+s\sum_{a>1}T^{aa}_1-\frac{2s}{3}\sum_{a,b}T^{ab}_0T^{ba}_0+
$$
$$
+\Delta_0\Delta_1-\Delta_1-\frac{1}{3}\Delta_0^3+\frac{2}{3}\Delta_0^2+\frac{(s
^2-1)}{3}\Delta_0.
$$
The Hamiltonians can be expressed in terms of $\Delta_i$:
$$
\frac{H_{1}}{\beta}=-\Delta_1+\frac{1}{2}\Delta_0^2-\frac{1}{2}\Delta_0,
$$
$$
\frac{H_{2}}{\beta^2}=\Delta_2-\Delta_0\Delta_1+\Delta_1+\frac{1}{3}\Delta_0^3-\frac{1}{2}\Delta_0^2+\frac{1}{6}\Delta_0.
$$
\section*{Acknowlegements} 
The authors thank M.L.Nazarov for valuable remarks and discussions, and I.Marshall and A.Schwarz for the careful proofreading.  
The research by M.M. was supported in part by the Simons Foundation and
RFBR grant 16-01-00562.
S.K appreciates the support of RSF grant, project 16-11-10316 dated
11.05.2016.

\end{document}